\documentclass{article}

\usepackage{fullpage}
\usepackage{textcomp}
\usepackage{amsmath}
\usepackage{amsfonts}
\usepackage[utf8]{inputenc}
\usepackage[square]{natbib}
\usepackage{graphicx}
\usepackage{mathrsfs}
\usepackage[labelfont=bf,textfont=sl,tableposition=top]{caption}
\usepackage{multirow}
\usepackage{setspace}
\usepackage{url}
\usepackage[ruled]{algorithm2e}
\usepackage{amsthm}

\theoremstyle{plain}

\theoremstyle{definition}
\newtheorem{exmp}{Example}

\def\Dto{\stackrel{d}{\longrightarrow}}
\title{Bayesian Inference from Composite Likelihoods, with an
  Application to Spatial Extremes}
\author{Daniel Cooley$^\dag$ \and Anthony C. Davison$^\ddag$ \and
  Mathieu Ribatet$^{\S,}$\thanks{Corresponding author. Email:
    mathieu.ribatet@math.univ-montp2.fr. Phone: +33 (0)4 67 14 41 98}}

\begin{document}
\maketitle
\begin{center}
  $^\dag$ Department of Statistics, Colorado State University

  $^\ddag$ Institute of Mathematics, \'Ecole Polytechnique F\'ed\'erale de
  Lausanne

  $^\S$ Department of Mathematics, Universit\'e Montpellier II
\end{center}

\begin{abstract}
  \noindent
  Composite likelihoods are increasingly used in applications where
  the full likelihood is analytically unknown or computationally
  prohibitive. Although some frequentist properties of the maximum composite likelihood estimator are akin to those of the maximum likelihood
  estimator, Bayesian inference based on composite likelihoods is
  in its early stages. This paper discusses inference when one uses
  composite likelihood in Bayes' formula.  We establish that using a
 composite likelihood results in a proper posterior density, 
  though it can differ considerably from that stemming 
  from the full likelihood.
%  The question is then how one can perform appropriate 
%  Bayesian inference when using a composite likelihood.
  %though not the usual one.  
  Building on previous work on composite
  likelihood ratio tests, we use asymptotic theory for misspecified
  models to propose two adjustments to the composite likelihood to 
  obtain %credible intervals with reasonable coverage.
  appropriate inference.
  We also investigate use of the Metropolis Hastings algorithm and 
  two implementations of the Gibbs sampler for 
  obtaining draws from the composite posterior.
%  how these adjustments can be applied in 
%  Markov chain Monte Carlo algorithms to
%  obtain realistic credible intervals. 
  We test the methods on
  simulated data and apply them to a spatial extreme rainfall dataset.
  For the simulated data, we find that posterior credible intervals 
  yield appropriate empirical coverage rates.
  For the extreme precipitation data, we are able to
  both effectively model marginal behavior throughout 
  the study region and obtain appropriate measures of spatial 
  dependence.

  \bigskip
  \noindent \textbf{Keywords:} Bayesian hierarchical model; Composite
  likelihood; Gibbs sampler; Markov chain Monte Carlo; Max-stable process; Metropolis--Hastings
  algorithm; Posterior coverage; Rainfall data.
\end{abstract}

\section{Introduction}
\label{sec:introduction}

\subsection{Motivation}

The likelihood function is central to both frequentist and Bayesian
inference, but in many modern settings it may be infeasible to
calculate it, either because no analytical form is available, or
because such a form is known but is computationally prohibitive. The
first difficulty arises with max-stable processes, which are used to
construct probability models for complex rare events, but for which
closed forms are typically available only for the bivariate marginal
densities \citep{Smith1990,Schlather2002,deHaan2006,Kabluchko2009},
though \citet{Genton.Ma.Sang:2011} show that substantial efficiency
gains are possible if trivariate margins can be used. The second
difficulty may be experienced when dealing with Gaussian random fields
on large lattices. Both of these problems and many other similar ones
can be tackled using composite likelihoods. \citet{Padoan2010} and
\citet{Gholamrezaee2010} propose the use of composite likelihood based
on marginal events to fit max-stable processes, and \citet{Rue2002}
have used composite likelihoods based on omitting components of the
full likelihood in approximating Gaussian random fields.
\citet{Ryden1998} describe the use of pseudo-likelihood, a form of
composite likelihood, in simulation-based inference involving missing
data, and show that their approach leads to a valid Markov chain
simulation algorithm.

Frequentist methods for composite likelihoods have been used for some
time (for an overview, see \citet{Varin2008}), but little work has
been done to explore how composite likelihoods could be employed in a
Bayesian framework.  The motivating application for this work is the
spatial modelling of extremes.  Recently authors (e.g.,
\citet{Padoan2010} and \citet{Gholamrezaee2010}) have used
composite likelihoods to fit max-stable models, enabling the
researchers to successfully model dependence between observations.
However, the frequentist methods they employ may not be flexible
enough
to accurately fit the marginal behavior across the study region. 
%to fully capture the marginal behavior.  Other authors (e.g,
\citet{Cooley2007} and \citet{Sang2009} have used Bayesian
hierarchical spatial models to capture the marginal effects for
spatial extremes, but have not used the max-stable process models
suggested by extreme value theory to describe the dependence in the
data.  The goal of this work is combine these two approaches,
and this entails appropriately deploying a composite likelihood within
a Bayesian framework.

\subsection{Likelihood asymptotics for composite likelihoods}

Although it has numerous antecedents, the notion of a composite
likelihood was crystallized by \citet{Lindsay1988}, who defined it as
a combination of valid likelihood entities.  Consider a random vector
$Y \in \mathbb{R}^K$ with probability density function $f(y; \theta)$
where $\theta \in \mathbb{R}^p$ is an unknown parameter vector.  Let
$\{\mathscr{A}_i: i \in I \}$, $I \subset \mathbb{N}$, be a set of
marginal or conditional events for $Y$ and let $\{w_i, i \in I\}$ be a
set of non-negative weights.  A composite likelihood is defined as
\begin{equation}
  \label{eq:clik}
  L_c(\theta;y) = \prod_{i \in I} f(y \in \mathscr{A}_i;\theta)^{w_i},
\end{equation}
with corresponding log-composite likelihood 
\begin{equation}
  \label{eq:lclik}
  \ell_c(\theta;y) = \sum_{i \in I} w_i \log f(y \in
  \mathscr{A}_i;\theta).
\end{equation}

Below  we assume that $n$ independent replicates $Y^1,\ldots, Y^n$ of $Y$ are available, yielding a total composite likelihood and log likelihood of the form
$$
 L^{\rm tot}_c(\theta;y) = \prod_{j=1}^n \prod_{i \in I} f(y^j \in \mathscr{A}_i;\theta)^{w_i},\quad 
  \ell^{\rm tot}_c(\theta;y) = \sum_{j=1}^n\sum_{i \in I} w_i \log f(y^j \in  \mathscr{A}_i;\theta), 
$$
and consider asymptotics as $n\to\infty$, with a fixed number of observations $K$ in each replicate.  
The development below is simpler if we work with quantities that remain of order one as $n\to\infty$, and we shall do so wherever possible.

If the true likelihood is unavailable or difficult to work with,
$\theta$ is often estimated by the maximum composite likelihood
estimator $\hat \theta_c$.  Let $\theta_0$ denote the true value of
the parameter.  As each term on the right-hand side of
equation~\eqref{eq:lclik} is a valid loglikelihood, the composite
score function $\nabla \ell^{\rm tot}_c(\theta;y)$ %obtained by differentiating
%it 
is a linear combination of unbiased estimating functions and so has
mean zero. Under appropriate regularity conditions, therefore, the
maximum composite likelihood estimator $\hat{\theta}_c$ converges in distribution as follows,
\begin{equation}
  \label{eq:asympMCLE}
  \sqrt{n} \{ H(\theta_0) J(\theta_0)^{-1} H(\theta_0)
  \}^{1/2} (\hat{\theta}_c - \theta_0 ) \stackrel{d}{\longrightarrow}
  N\left(0, \mbox{Id}_p \right), \qquad n \rightarrow \infty, 
\end{equation}
where $M^{1/2}$ denotes a matrix square root, i.e.,\ $\{M^{1/2}\}^T
M^{1/2} = M$, $\mbox{Id}_p$ denotes the $p \times p$ identity matrix,
$H(\theta_0) = -\mathbb{E}[\nabla^2 \ell_c(\theta_0;Y)]$ and
$J(\theta_0) = \mbox{Var}[\nabla \ell_c(\theta_0;Y)]$, where the
expectations are with respect to the full density. Both $H(\theta_0)$ and $J(\theta_0)$ are positive definite in a regular model, and 
both are of order one as $n\to\infty$.

Essentially the usual regularity conditions for the asymptotic
normality of the maximum likelihood estimator as $n\to\infty$ apply 
\citep[Sec.~4.4.2]{Davison2003}, but the parameter $\theta$ must
be identifiable from the densities appearing in~(\ref{eq:lclik}). The
limiting distribution in equation~(\ref{eq:asympMCLE}) also stems from
the behavior of the maximum likelihood estimator under
mis-specification \citep{Kent1982}.  The maximum composite likelihood
estimator may thus be viewed as resulting from a mis-specified, or,
more accurately, under-specified, statistical model, leading to
consistent estimation but with a ``sandwich'' variance estimator of
the type arising in longitudinal data analysis and many other domains.

\subsection{Bayesian inference with a composite likelihood}
\label{sect:1.3}

Bayesian inference based on composite likelihoods has been little explored.
%So far as we are aware, the only previous paper to employ a composite
%likelihood within a Bayesian framework is 
Motivated by the spatial extremes problem mentioned above,
\citet{Smith2009} use a
pairwise likelihood and Markov chain Monte Carlo simulation to fit a
max-stable model for rainfall at five sites in South-West England.
They obtain a posterior by replacing the unavailable full likelihood with the 
pairwise likelihood, but although they mention that this substitution may lead to overly precise inferences,
they do not describe how to correct this.
%We demonstrate this effect later.
%they do not provide ways to overcome this drawback.  
\citet{Pauli2011} independently suggest 
the adjustment to the composite likelihood called by us the
magnitude adjustment in Section \ref{sec:magnitude-adjustment}, 
 establish the asymptotic normality of the corresponding composite
posterior and apply the method to a five-dimensional 
data set on air pollution.

Related to Bayesian inference with composite likelihoods is work in
Bayesian methods when one lacks or wishes to avoid using the true
likelihood.
\citet{Monahan1992} explore the validity of a posterior when the likelihood is not the 
conditional density of the data given the parameter, and propose both an alternative definition based
on the coverage of posterior sets and a test that can be used to
invalidate a particular replacement likelihood.
\citet{Lazar2003} applies this test when an empirical likelihood is
used in place of a parametric one.  Other work on Bayesian methods
with conditional or pseudo likelihoods (e.g.,
\citet{Efron1993}, \citet{Chang2006} and \citet{Ventura2009}) is typically motivated by a
desire to avoid specifying a full likelihood when there are nuisance
parameters, and thus focuses on Bayesian implementation using a
pseudo-likelihood, which is often a marginal, conditional or profile
likelihood for the parameters of interest.  Like
\citeauthor{Monahan1992}, our ultimate aim is the practical one of
using a composite likelihood to provide valid inferences; for example, the resulting posterior confidence sets should be correctly calibrated. 

Provided that $\int L^{\rm tot}_c(\theta;y) \pi(\theta) \mbox{d$\theta$} $ is
finite, we use (\ref{eq:clik}) to define a composite posterior density as
\begin{equation}
  \label{eq:compPosterior}
  \pi_c(\theta \mid y) = \frac{L^{\rm tot}_c(\theta;y) \pi(\theta)}{\int
    L^{\rm tot}_c(\theta;y) \pi(\theta) \mbox{d$\theta$}},
\end{equation}
where $\pi(\cdot)$ is the prior density.  The first question arising
is under what circumstances $\int L^{\rm tot}_c(\theta;y) \pi(\theta)
\mbox{d$\theta$} < \infty$, so that (\ref{eq:compPosterior}) is
well-defined.  In Bayesian analysis, integrability questions usually
arise when discussing improper priors; but here we suppose that $\pi(\cdot)$ is proper.  Then a sufficient condition 
for (\ref{eq:compPosterior}) to be proper is that for
each $i$ there exists a finite $b_i$ such that $\sup_\theta f(y \in
\mathscr{A}_i;\theta)\leq b_i$, since in that case
\begin{equation}
  \label{eq:validCompPost}
  \int L^{\rm tot}_c(\theta;y) \pi(\theta) \mbox{d$\theta$} = \int \prod_j\prod_{i \in
    I} f(y^j \in \mathscr{A}_i;\theta)^{w_i} \pi(\theta)
  \mbox{d$\theta$} \leq \prod_{i \in I} b_i^{nw_i} < \infty.
\end{equation}
The boundedness of $f(y\in \mathscr{A}_i;\theta)$ holds in many cases,
and in cases of doubt it can be imposed by recalling that in practice
continuous observations are always rounded to some extent.  The
correct likelihood is therefore a product of probabilities obtained as
differences of cumulative distribution functions, for which $b_i\equiv
1$. The rounding is often ignored so that simpler density function
approximations to the correct likelihood may be used, but if these
approximations lead to difficulties, then we may choose to work with
the correct likelihood; see, e.g., \citet{Copas:1972}. As this
rounding argument applies to any probability elements in
\eqref{eq:clik}, and, with minor changes, also applies to the modified
composite likelihoods used below, in practice we may always arrange
that $\int L^{\rm tot}_c(\theta;y)\pi(\theta) \mbox{d$\theta$} <
\infty$ and thus that (\ref{eq:compPosterior}) is proper.

% \textbf{Mathieu: What we really need is that each $f(y \in
%   \mathscr{A}_i;\theta)$ is upper bounded $\pi(\theta)$--almost
%   surely.}

Since the composite likelihood is not the likelihood believed to have
generated the data, the naive implementation of a composite posterior
may give misleading inferences, as we now illustrate.

\begin{exmp}%[Gaussian process with known scale parameter]
  \label{exmp:GaussianProcess}
  Let $\{Y(x)\}$ be a stationary Gaussian process with unknown mean
  $\mu \in \mathbb{R}$ and with covariance function $\gamma(h) = \tau
  \exp(-h / \omega)$, where the sill $\tau > 0$ is unknown but the 
  scale $\omega > 0$ is known. Let $\{y(x_1), \ldots, y(x_K)\}$ be one
  realisation of this process at locations $x_1, \ldots, x_K \in
  \mathbb{R}$. Now consider a prior density of the form
  $\pi(\theta) = \pi(\mu) \pi(\tau)$, where $\pi(\mu) \sim N(a,
  b)$ and $\pi(\tau) \sim {\rm IG}(c, d)$, i.e., an inverse Gamma
  distribution with shape $c$ and scale $d$.
 
 Here the prior densities are conjugate for $\pi(\theta
  \mid y)$, so the full conditional distributions needed for Gibbs sampling are easily found to be
  \begin{equation*}
    \pi(\mu \mid \cdots) \sim N\left(\tilde{\mu},
      \tilde{\sigma}^2\right), %\quad\mbox{and}
      \qquad  \pi(\tau \mid
    \cdots) \sim {\rm IG}\left\{c + \frac{K}{2}, d + \frac{1}{2} (y -
      \mu \mathbf{1})^T \Sigma^{-1} (y - \mu \mathbf{1})\right\},
  \end{equation*}
  where $\tilde{\sigma}^2 = \left(b^{-1} + \tau^{-1} \mathbf{1}^T
    \Sigma^{-1} \mathbf{1} \right)^{-1}$, $\tilde{\mu} =
  \tilde{\sigma}^2 \left(a b^{-1} + \tau^{-1} \mathbf{1}^T \Sigma^{-1}
    y\right)$ and $\Sigma$ is the correlation matrix derived from
  $\gamma(\cdot)$.

The full conditional pairwise distributions are also readily available, and are
%   \begin{align*}
%     \pi_p(\mu \mid \cdots) &\sim N\left(\tilde{\mu}_p,
%       \tilde{\sigma}_p^2\right)\\
%     \pi_p(\tau \mid \cdots) &\sim {\rm IG}\left\{c + \frac{k(k-1)}{2},
%       d + \frac{1}{2} \sum_{i< j} (y_{ij} - \mu \mathbf{1})^T
%       \Sigma_{ij}^{-1} (y_{ij} - \mu \mathbf{1}) \right\}\\
%     &\sim {\rm IG}\left\{c + \frac{k(k-1)}{2}, d + \frac{1}{2}
%       (y_p - \mu \mathbf{1})^T \Sigma_p^{-1} (y_p - \mu
%       \mathbf{1}) \right\},
%   \end{align*}
  \begin{equation*}
    \pi_p(\mu \mid \cdots) \sim N\left(\tilde{\mu}_p,
      \tilde{\sigma}_p^2\right), \qquad \pi_p(\tau
    \mid \cdots) \sim {\rm IG}\left\{c + \frac{K(K-1)}{2}, d +
      \frac{1}{2} (y_p - \mu \mathbf{1})^T \Sigma_p^{-1}
      (y_p - \mu \mathbf{1}) \right\},
  \end{equation*}
  where $\tilde{\sigma}_p^2 = \left(b^{-1} + \tau^{-1} \mathbf{1}^T
    \Sigma_p^{-1} \mathbf{1}\right)^{-1}$, $\tilde{\mu}_p =
  \tilde{\sigma}_p^2 \left(a b^{-1} + \tau^{-1} \mathbf{1}^T
    \Sigma_p^{-1} y_p\right)$, $\Sigma_p$ is a block diagonal
  matrix with blocks 
  \begin{equation*}
    \begin{bmatrix}
      1 & \tau^{-1} \gamma(x_i - x_j)\\
      \tau^{-1} \gamma(x_i - x_j) & 1
    \end{bmatrix},
    \qquad 1 \leq i < j \leq K,
  \end{equation*}
  and $y_p = (y_1, y_2, y_1, y_3, \ldots, y_1, y_K, y_2, y_3, \ldots,
  y_2, y_K, \ldots, y_{K-1}, y_K)^T$.\hfill $\square$
\end{exmp}

Example~\ref{exmp:GaussianProcess} shows that, as might be
expected, the full conditional densities derived from the pairwise
likelihood differ from those derived from the
full likelihood. Since $\mathbf{1}^T A \mathbf{1}$ is the sum of
all entries of the matrix $A$ and $\Sigma_p$ is block diagonal, it is
not difficult to show that
\begin{equation*}
  \mathbf{1}^T \Sigma_p^{-1} \mathbf{1} = 2 \sum_{i=1}^{K-1}
  \sum_{j=i+1}^K \left\{1 + \tau^{-1} \gamma(x_i - x_j) \right\}^{-1}
  \geq \frac{\tau K (K-1)}{1 + \tau}, \qquad 
  \mathbf{1}^T \Sigma^{-1} \mathbf{1} \leq K,
\end{equation*}
so, in particular, 
\begin{equation*}
  \frac{\tilde{\sigma}_p^2}{\tilde{\sigma}^2} \leq \frac{(1 +
    \tau)(\tau + b K)}{(1 + \tau) \tau + b \tau K (K - 1)},
\end{equation*}
and when $\tau$ is fixed, $\tilde{\sigma}_p^2 / \tilde{\sigma}^2
\downarrow 0$ as $K \to \infty$.

\begin{figure}
  \centering
  \includegraphics[width=0.75\textwidth]{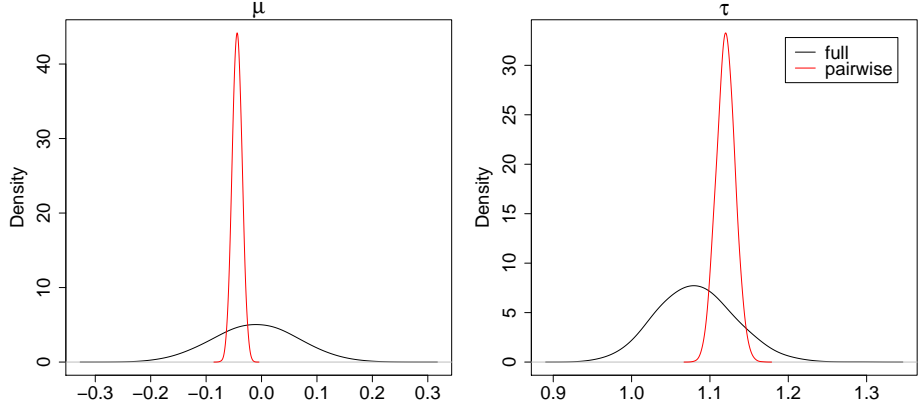}
  \caption{Marginal full and pairwise posterior densities for the mean
    $\mu$ (left) and sill $\tau$ (right), derived from $n=50$
    realisations of a Gaussian process observed at $K=20$ locations
    having an exponential covariance function with $\mu=0$, $\tau = 1$
    and $\omega = 3$.}
  \label{fig:fullPairPostDist}
\end{figure}

%To emphasize a bit more on this, 
To illustrate this discussion, Figure~\ref{fig:fullPairPostDist}
shows posterior marginal density estimates for $\mu$ and $\tau$ based on the composite and full likelihoods,  found 
using a Gibbs sampler. These densities were
obtained by taking the same setting as in
Example~\ref{exmp:GaussianProcess} with $\mu = 0$, $\tau = 1$ and
$\omega = 3$, the last taken as constant in the sampling algorithm, with $K=20$, and with the locations $x_1,\ldots, x_K$ taken uniformly at random in $[0,20]$.  There are $n=50$ independent replicates of these data. 
A Gaussian prior with mean $0$
and variance $100$ was placed on $\mu$, and independently an inverse gamma prior with
shape $1/10$ and scale $1$ was placed on $\tau$. The
marginal composite posterior densities are much too concentrated,
because the pairwise likelihood treats the pairs of observations as though they were mutually independent and thus uses each observation repeatedly---see the definition of $y_p$ in Example~\ref{exmp:GaussianProcess}.

% To illustrate this, we consider a one-dimensional stationary Gaussian
% process $Y(\cdot)$ with mean $\mu \in \mathbb{R}$ and an isotropic
% exponential covariance function $\gamma(h) = \tau \exp(-h/\omega)$,
% where $h$ denotes the distance between locations and both the sill and
% range parameters $\tau$ and $\omega$ are positive. Since the full
% likelihood is known and the pairwise densities are readily available,
% we can easily compare the naive posterior density based on the
% pairwise likelihood with the true posterior.  We simulate $n=50$
% replications of a Gaussian process with   Since the full
% likelihood is known, we may compare the posterior density associated
% with the full likelihood, which is proportional to $L(\theta; y)
% \pi(\theta)$, to a naively-implemented composite likelihood, which is
% proportional to $L_c(\theta; y)\pi(\theta)$.  Here, we use a pairwise
% likelihood,
% \begin{equation*}
%   L_c(\theta; y) = \prod_{m=1}^n \prod_{i=1}^{K-1} \prod_{j=i+1}^K
%   f(y_{m,i}, y_{m,j}; \theta),
% \end{equation*}
% where $y_{m,i}$ denotes the observation at location $i$ for
% realization $m$. Figure~\ref{fig:fullPairPostDist} plots kernel
% density estimates of the posterior density functions. As expected, the
% marginal composite posterior densities are much too concentrated,
% because the pairwise likelihood uses each observation repeatedly.

The aim of this paper is to propose a framework for approximate
Bayesian inference from composite likelihoods when the full likelihood
is not available. Our aim is to obtain composite posterior
distributions that give credible intervals with reasonable coverage. Section~\ref{sec:adjustm-comp-likel} introduces two
adjustments to the composite likelihood that are intended to retrieve
some of the desirable properties given by the usual likelihood.
Section~\ref{sec:mcmc-samplers} shows how these adjustments can be
incorporated into Markov chain Monte Carlo samplers, and their
performance in simulation studies is discussed in
Section~\ref{sec:simulation-study}. Section~\ref{sec:application}
gives a case study on the modelling of extreme rainfall around Zurich.  The paper closes with a brief discussion and two technical appendices.

\section{Adjustment of the composite likelihood}
\label{sec:adjustm-comp-likel}

We ultimately wish to perform a Bayesian analysis, in which setting there is no
``true'' parameter value $\theta_0$. However, we use asymptotic
relationships developed under the frequentist paradigm to 
adjust the likelihood to obtain appropriate inference
for the composite posterior, 
%adjust the likelihood, 
and thus speak of $\theta_0$ throughout this section.

The theory of unbiased estimating functions applied to the score
functions of composite likelihood implies that under suitable
regularity conditions, the modes of a composite posterior and of the
full posterior density will approach one another as the sample size
increases; see Figure~\ref{fig:fullPairPostDist}. However, the
figure also shows that the composite posterior density can differ
significantly in spread
from the true one, because the composite likelihood treats the 
events $\{\mathscr{A}_i, i \in I\}$ as though they were mutually independent. Below we seek to modify the composite
likelihood in order to mitigate this.

Suppose that the parameter $\theta = (\phi^T, \psi^T)^T$ has true
value $\theta_0=(\phi_0^T,\psi_0^T)^T$ and that $\psi$ contains $q$
elements. Let $\tilde{\theta}$ be the restricted maximum likelihood
estimator, obtained by maximizing the full log likelihood $\ell(\theta;Y)$ over $\theta$
with $\psi$ held fixed at $\psi_0$ and let $\tilde{\theta}_c$ be the
restricted maximum composite likelihood estimator, which 
maximizes~(\ref{eq:clik}) with $\psi$ held fixed at
$\psi_0$. Then as $n \to \infty$,
\begin{equation}
  \label{eq:likRatio}
  \Lambda(\psi_0) = 2 \{\ell(\hat{\theta};Y) -
  \ell(\tilde{\theta};Y)\} \stackrel{d}{\longrightarrow} \chi^2_q
\end{equation}
whereas for the composite likelihood, 
\begin{equation}
  \label{eq:likRatioMiss}
  \Lambda_c(\psi_0) = 2 \{\ell_c(\hat{\theta}_c;Y) -
  \ell_c(\tilde{\theta}_c;Y) \} \stackrel{d}{\longrightarrow}
  \sum_{i=1}^q \lambda_i X_i 
\end{equation}
where $X_1,\ldots,X_q$ are independent
$\chi_1^2$ random variables, $\lambda_1,\ldots,\lambda_q$ are the eigenvalues of the $q\times q$ matrix
$\{H(\theta_0)^{-1}J(\theta_0) H(\theta_0)^{-1}\}_\psi
[\{H(\theta_0)^{-1}\}_\psi ]^{-1}$, and $A_\psi$ denotes the sub-matrix of a
matrix $A$ corresponding to the elements of $\psi$ \citep{Kent1982}.
These relationships have previously been exploited to provide
likelihood ratio tests \citep{Rotnitzky1990, Chandler2007} suitable
for misspecified models. Here we aim to recover convergence in
distribution to the usual $\chi^2$ distribution through two different
modifications of the composite likelihood: a magnitude adjustment and
a curvature adjustment. The reasons for such modifications is to make
the composite likelihood ratio, which appears in the Metropolis--Hastings
algorithm but is hidden in the Gibbs sampler, behave in distribution as
it would if a full likelihood were available. In the remainder of
this section we consider only the case where $\psi$
has dimension zero, but in
Section~\ref{sec:adaptive-gibbs}  we will show how partitioning
$\theta$ can yield better coverage.

\subsection{Magnitude adjustment}
\label{sec:magnitude-adjustment}

The magnitude adjustment to the composite log likelihood is inspired
by \citet{Rotnitzky1990}, who, in the context of hypothesis testing in
longitudinal studies, estimate $\lambda_1, \ldots, \lambda_q$
from estimates of $H(\theta_0)$ and $J(\theta_0)$,
and use them to calculate the appropriate rejection region for the
$\chi^2$ test based on (\ref{eq:likRatioMiss}).

We define the magnitude adjustment by
\begin{equation}
  \label{eq:adjCompLikRJ}
  \ell_{\rm magn}(\theta;y) = k \ell^{\rm tot}_c(\theta;y), \qquad \theta \in
  \Theta,
\end{equation}
where $k$ is a positive constant; (\ref{eq:adjCompLikRJ}) was also suggested by \cite{Pauli2011}.
With this modification and as $n\to\infty$ we have
\begin{equation}
  \label{eq:asympAdjCompLikRJ}
  \Lambda_{\rm magn}(\psi_0) = 2 \{\ell_{\rm magn}(\hat{\theta}_c;Y) -
  \ell_{\rm magn}(\tilde{\theta}_c;Y) \} \stackrel{d}{\longrightarrow}
  k \sum_{i=1}^q \lambda_i X_i
\end{equation}
and 
\begin{equation*}
  \mathbb{E}[\Lambda_{\rm magn}(\psi_0)] \longrightarrow k \sum_{i=1}^q
  \lambda_i ,  \qquad \mbox{Var}[\Lambda_{\rm
    magn}(\psi_0)] \longrightarrow 2 k^2 \sum_{i=1}^q \lambda_i^2.
\end{equation*}
Setting $k = q / \sum_{i=1}^q \lambda_i$ therefore ensures that
$\mathbb{E}[\Lambda_{\rm magn}(\psi_0)]$ converges to
$\mathbb{E}[\chi^2_q]= q$, but the higher moments of (\ref{eq:asympAdjCompLikRJ}) will not match those of $\chi^2_q$ 
unless all the $\lambda_i$'s are equal or $q = 1$. For our purposes, we consider
the case where $\phi$ has dimension zero, i.e.,\ $k = p / \sum_{i=1}^p
\lambda_i$ where $\lambda_1, \ldots, \lambda_p$ are the eigenvalues of
$H(\theta_0)^{-1} J(\theta_0)$. %A Satterthwaite adjustment
\citet{Varin2008} proposes a Satterthwaite 
adjustment to match the first two moments
of $\Lambda_{\rm magn}(\psi_0)$ and $\chi^2_q$, though their higher moments
would still differ.

\subsection{Curvature adjustment}
\label{sec:curvature-adjustment}

Another strategy is to modify the curvature of the composite
likelihood around its global maximum $\hat{\theta}_c$ by considering the adjustment given by 
\begin{equation}
  \label{eq:adjlclik}
  \ell_{\rm curv}(\theta;y) = \ell^{\rm tot}_c(\theta^*;y), \qquad  \theta^* =
  \hat{\theta}_c + C (\theta - \hat{\theta}_c),
\end{equation}
for some constant $p \times p$ matrix $C$. Clearly $\hat{\theta}_c$ is also a
global maximum for $\ell_{\rm curv}$, and
\begin{equation*}
  \nabla \ell_{\rm curv}(\theta;y) = C^T \nabla \ell^{\rm tot}_c(\theta; y)
  \arrowvert_{\theta = \theta^*}, \qquad \nabla^2 \ell_{\rm
    curv}(\theta; y) = C^T \nabla^2 \ell^{\rm tot}_c(\theta ; y)
  \arrowvert_{\theta = \theta^*} C.
\end{equation*}

Under mild conditions, Taylor expansion of the usual log-likelihood and the
asymptotic normality of the maximum likelihood estimator $\hat{\theta}$ yield convergence of the
likelihood ratio statistic in distribution to a $\chi^2$ variable
\citep[Sec. 4.5]{Davison2003}. More precisely, the facts that \begin{equation*}
  \Lambda(\theta_0) \Dto n (\hat{\theta} - \theta_0)^T \Sigma
  (\hat{\theta} - \theta_0), \quad n\to\infty, 
\end{equation*}
for some $q \times q$ covariance matrix $\Sigma$ depending only on
$\mathbb{E}[\nabla^2 \ell(\theta_0;Y)]$ and
\begin{equation*}
  \sqrt{n} \Sigma^{1/2} (\hat{\theta} - \theta_0) \Dto N(0,
  \mbox{Id}_p), \quad n\to\infty, 
\end{equation*}
ensure that $\Lambda(\theta_0)$ converges in distribution to a
$\chi^2_p$ variable. This occurs because $-n^{-1} \nabla^2
\ell(\hat{\theta}; y)$ converges almost surely to the rescaled inverse of the
asymptotic covariance matrix of the maximum likelihood
estimator, the Fisher information  in a single $Y$.

This suggests that we should try to ensure that $- n^{-1} \nabla^2
\ell_{\rm curv}(\hat{\theta}_c; y)$ converges almost surely to the
inverse of the asymptotic covariance matrix of the maximum
composite likelihood estimator, i.e.,\ $H(\theta_0) J(\theta_0)^{-1}
H(\theta_0)$, by taking any semi-definite negative matrix $C$ such
that
\begin{equation}
  \label{eq:mtxCDefn}
  C^T H(\theta_0) C = H(\theta_0) J(\theta_0)^{-1} H(\theta_0).
\end{equation}
One possible choice, $C = M^{-1} M_A$, where $M_A^T M_A^{\phantom{T}}
= H(\theta_0) J(\theta_0)^{-1} H(\theta_0)$ and $M^T M = H(\theta_0)$,
corresponds to a suggestion of \citet{Chandler2007} for hypothesis
testing for clustered data using the independence
log-likelihood. However, the matrix square roots $M$ and $M_A$ are not
unique, and although the choice is immaterial for composite
likelihoods that are quadratic in the neighborhood of
$\hat{\theta}_c$, it might be necessary to ensure that the mapping
(\ref{eq:adjlclik}) preserves any directions of asymmetry. For this
reason we use singular value decompositions for $M$ and $M_A$ for the
curvature adjustments in this paper.

\subsection{Properties of the adjustments}
\label{sec:prop-adjustm}

Although both adjustments rely on the idea of recovering the usual
convergence to a $\chi^2$ variable, they express different aspects of
this.  The magnitude adjustment (\ref{eq:adjCompLikRJ}) is an ``overall'' adjustment, intended to scale the composite likelihood down to the appropriate
magnitude; in Figure~\ref{fig:fullPairPostDist} it amounts to raising the narrower curve to a power and thus giving a nonlinear transformation of the vertical axis. Therefore all (local) extrema are left unchanged, because
$\nabla \ell_{\rm magn}(\theta ; y) = 0$ implies that $ \nabla
\ell^{\rm tot}_c(\theta; y) = 0$, and the composite and full posterior modes
should be approximately the same, because the composite score function
has mean zero. The curvature adjustment (\ref{eq:adjlclik}), on the other hand, stretches the horizontal axis linearly so that the curvature of $\ell_{\rm curv}(\theta;y) $ at $\hat\theta_c$ matches that of the large-sample log-density of $\hat{\theta}_c$; thus this changes the locations of any local maxima other than the global maximum at $\hat{\theta}_c$. Therefore the magnitude adjustment might be more appropriate
if the full posterior distribution is multi-modal.

However only the curvature adjustment ensures that the convergence to
a $\chi^2$ distribution is met; the magnitude adjustment only gets the
first moment correct. This may have a strong impact on the shape of
the composite likelihood around $\hat{\theta}_c$, and therefore on the
composite posterior density.

\subsubsection{Asymptotic posterior distributions}
\label{sec:asympt-post-distr}

We now derive the asymptotic properties of the composite posterior distributions, both adjusted and 
unadjusted. 
Provided that the unadjusted
composite posterior is a valid distribution, it can be
shown under the usual regularity conditions that when $n$ is large enough (Appendix~\ref{sec:asympt-distr-post}), 
\begin{equation}
  \label{eq:asymptNonAdjPostDist}
  \pi_c(\theta \mid y) \stackrel{\cdot}{\sim} \mbox{N}\left\{\theta_0,
    n^{-1} H(\theta_0)^{-1} \right\}.
\end{equation}
Here and below we abuse notation; (\ref{eq:asymptNonAdjPostDist}) means that $\theta$ has the stated distribution, conditional on $y$, not that the posterior density has a distribution. 
Unlike in the usual case, the unadjusted composite posterior 
distribution does
not converge to the asymptotic distribution of the composite
likelihood estimator, given by~\eqref{eq:asympMCLE}.

In their investigation of the asymptotic distribution of the the
magnitude-adjusted posterior, \citet[page 8]{Pauli2011} state that the
posterior has approximately the correct variance ``by the $\chi^2$
approximation for the null distribution."  Further \citet[pages 8 \&
9]{Pauli2011} state that the approximation is asymptotically correct
when $p = 1$, and argue that the approximation represents an
improvement over the naive composite posterior when $p > 1$.  To
expand on this, as the scaling constant estimate $\hat k = p /
\sum_{i=1}^p \hat\lambda_i$ used for the magnitude adjustment
converges almost surely to $p / \mbox{tr}\{H(\theta_0)^{-1}
J(\theta_0) \}$ as $n \to \infty$, we conclude that (Appendix~\ref{sec:asympt-distr-post}),
\begin{equation}
  \label{eq:asymptMagnPostDist}
  \pi_{\rm magn}(\theta \mid y) \stackrel{\cdot}{\sim}
  \mbox{N}\left\{\theta_0, (np)^{-1} \mbox{tr}\{H(\theta_0)^{-1}
    J(\theta_0)\} H(\theta_0)^{-1} \right\}.
\end{equation}
Thus unless $\theta_0$ is scalar, i.e.,\ unless $p=1$, $\pi_{\rm magn}$
will differ from the asymptotic distribution given
by~\eqref{eq:asympMCLE}. Compared
to~\eqref{eq:asymptNonAdjPostDist}, the asymptotic variance is
inflated, because $\mbox{tr}\{H(\theta_0)^{-1} J(\theta_0) \} \geq p$; 
see Appendix~\ref{sec:asympt-vari-infl}.

Since the curvature adjustment %it is not difficult to see that
obtains the correct curvature, it is straightforward to see that
\begin{equation}
  \label{eq:asymptCurvPostDist}
  \pi_{\rm curv}(\theta \mid y) \stackrel{\cdot}{\sim}
  \mbox{N}\left\{\theta_0, n^{-1} H(\theta_0)^{-1} J(\theta_0)
    H(\theta_0)^{-1} \right\},
\end{equation}
which is exactly the asymptotic distribution of the maximum composite
likelihood estimator.

\subsubsection{Comparison of the Adjusted Likelihood to the Full Likelihood}
\begin{figure}
  \centering
  \includegraphics[angle = -90, width=2.5 in]{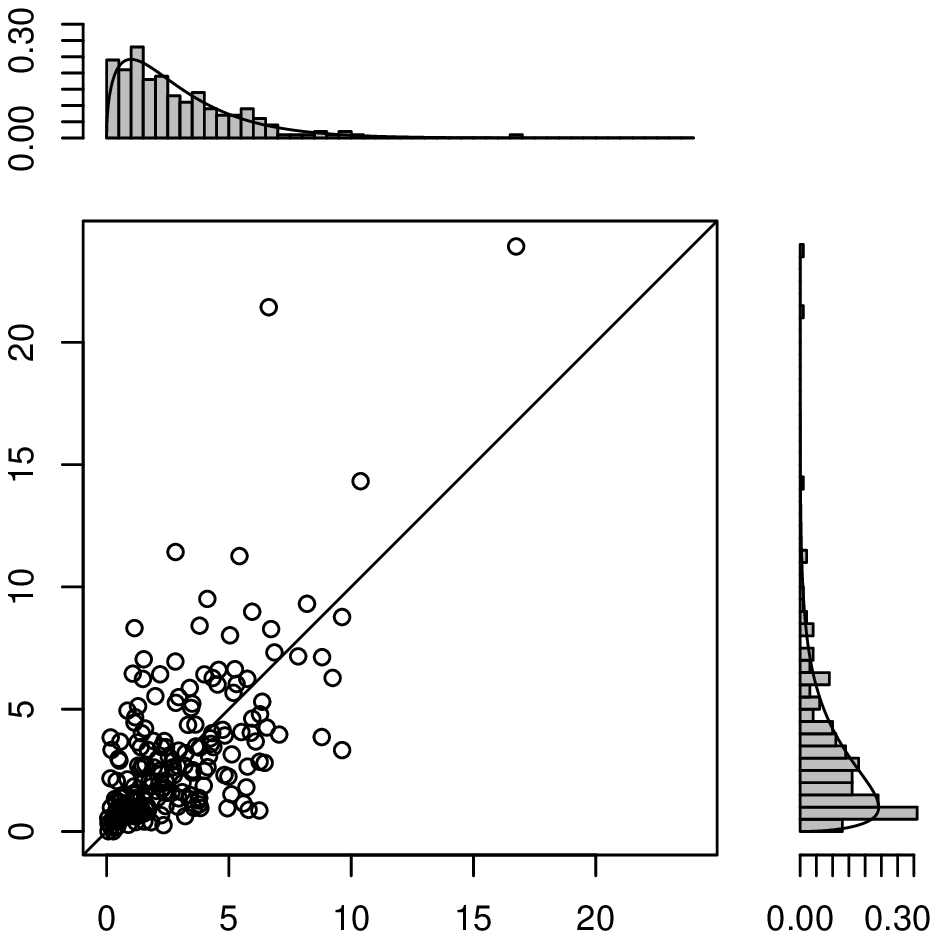}
  \includegraphics[angle = -90, width= 2.5 in]{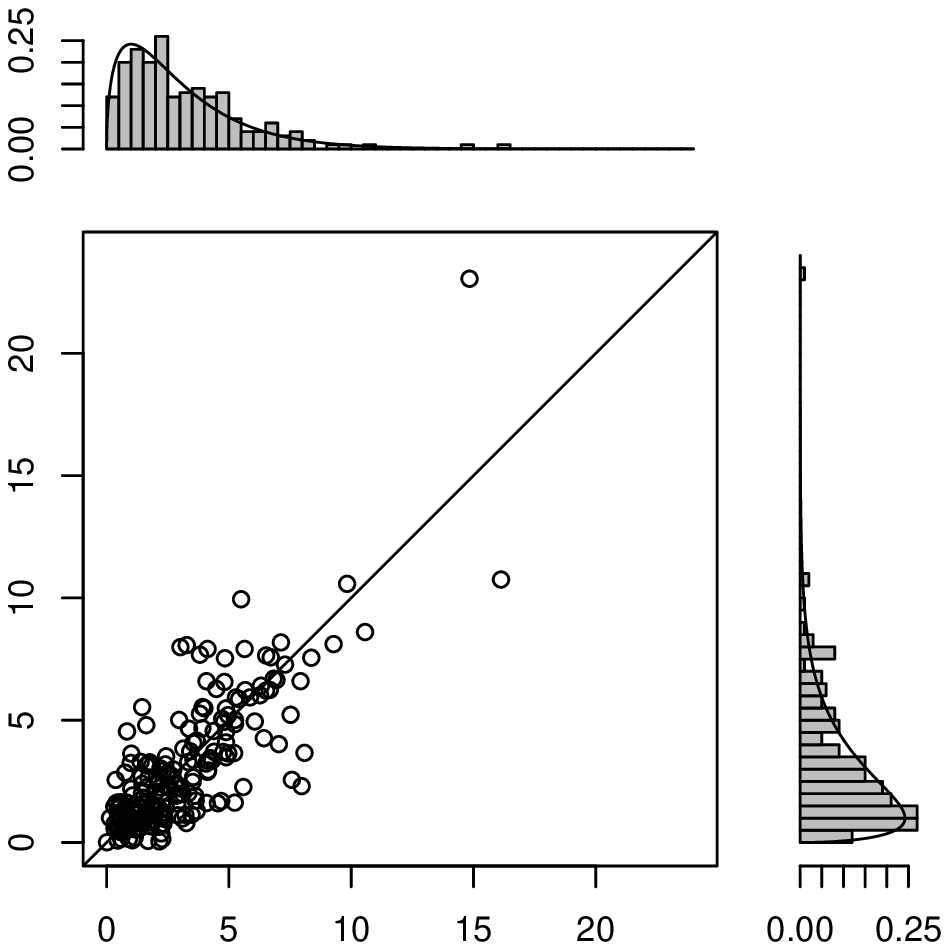}
  \caption{Comparison of 200 likelihood ratios for the Gaussian process
    simulation with $n$ replicates: $\Lambda_{\rm curv}(\theta_0)$ for the
    curvature-adjusted composite likelihood ($y$-axis) versus
    $\Lambda(\theta_0)$ ($x$-axis).  Left: $n=50$.  Right: $n=200$.}
  \label{fig:lhoodRatioSimulation}
\end{figure}

The magnitude or curvature adjustment will ensure only that the 
{\em distribution}\/ of the corresponding 
adjusted composite likelihood ratio, $\Lambda_{\rm adj}(\theta_0) = 2
\{ \ell_{\rm adj}(\hat \theta_c; y) - \ell_{\rm adj}(\theta_0; y) \}$, 
will approximate the $\chi^2_p$ distribution
of the true likelihood ratio, $\Lambda(\theta_0)$.
%; here  $\ell_{\rm adj}$ may 
%denote either adjusted log-composite
%likelihood.  
However, since the composite likelihood should contain
some of the information in the full likelihood, one would hope that
$\Lambda_{\rm adj}(\theta_0) \approx \Lambda(\theta_0)$, i.e.,\ that the values of these ratios should be related. %The left panel of
Figure~\ref{fig:lhoodRatioSimulation} compares values of $\Lambda_{\rm
  curv}(\theta_0)$ and $\Lambda(\theta_0)$ for 200 datasets simulated as described in \S\ref{sect:1.3}.  Their correlation is $\hat r = 0.64$ when the number of replicate Gaussian processes is $n = 50$, and $\hat r = 0.79$ 
when $n = 500$: reasonable correlations, but not overwhelming.

Our aim in adjusting the likelihood is not to approximate the true 
likelihood---and in turn, approximate the full posterior---but rather to
obtain appropriate inference from a composite posterior.
%Unlike in previous publications on modified likelihood ratios
%\citep{Rotnitzky1990,Chandler2007}, we need to approximate the true
%likelihood at parameter values other than the true value $\theta_0$.
If we did wish to approximate $\ell(\theta_1)$ at $\theta_1 \in
\Theta$, then 
% and, for illustrative purposes, let us use the
 it can be shown using the 
curvature-adjusted likelihood that $2\{\ell_{\rm
  curv}(\hat \theta_c) - \ell_{\rm curv}(\theta_1)\}
\stackrel{d}{\rightarrow} X^TX$, where $X \sim N( \{H(\theta_0)
J^{-1}(\theta_0) H(\theta_0) \}^{1/2} (\theta_1 - \theta_0),
\mbox{Id}_p )$, whereas $2\{ \ell(\hat \theta) - \ell(\theta_1)
\} \stackrel{d}{\rightarrow} Y^TY$, where $Y \sim N\{ I(\theta_0)^{1/2}
(\theta_1 - \theta_0), \mbox{Id}_p \}$ and $I(\theta_0)$ is the Fisher
information matrix based on the full likelihood.  Obviously, the
approximation will degrade as $(\theta_1 - \theta_0)$ grows. Since the true
likelihood and information about $I(\theta_0)$ would not be available in a realistic application,
it seems unclear how to improve the approximation to the true
likelihood away from $\theta_0$.  
%In the following sections we investigate whether the
%relationship between the adjusted likelihood and the true likelihood
%is strong enough to provide adequate inference.
Simply put, by not having 
the full likelihood available, we lose information.
%The adjustments presented in Sections \ref{sec:magnitude-adjustment} and
%\ref{sec:curvature-adjustment} provide procedures for appropriately 
%accounting for the information in the composite likelihood.
%\footnote{Dan:  Cut the sentence at the end of this paragraph per Anthony's earlier comment}

%\begin{figure}
%  \includegraphics[angle=-90,width=\textwidth]{lhoodRatioSimulation}
%  \caption{Comparison of likelihood ratios for the Gaussian process
%    simulation.  Left: plot of the likelihood ratio $\Lambda_{\rm
%      curv}(\theta_0)$ for the curvature-adjusted composite likelihood
%    ($y$-axis) versus $\Lambda(\theta_0)$ ($x$-axis) for 200 different
%    Gaussian process realizations.  Center: corresponding likelihood
%    ratios for Metropolis--Hastings moves from $\theta^{(p)} =
%    (0,1,2.5)$ to $\theta^{(t)} = (0, 1, 2.48)$.  Right: likelihood
%    ratios for the the reverse moves.}
%  \label{fig:lhoodRatioSimulation}
%\end{figure}

\section{Markov chain Monte Carlo samplers}
\label{sec:mcmc-samplers}

This section describes implementations of Markov chain Monte Carlo basing Bayesian inference on composite
likelihoods. One must take care to show that MCMC algorithms will converge to the correct target distributions,  
as composite likelihoods, adjusted or not, are not valid likelihoods. We describe
the adjusted Metropolis--Hastings algorithm and the Gibbs sampler in turn.

\subsection{Adjusted Metropolis--Hastings algorithm}
\label{sec:adjust-metr-hast}

In Section~2 we suggested two adjustments intended to
provide approximations to the full likelihood ratios. We now discuss
an adjusted Metropolis--Hastings algorithm, given in
Algorithm~\ref{alg:adjMH}, and verify that it has the desired stationary distribution.

\begin{algorithm}
%  \LinesNumbered
  \SetKwInOut{Input}{Input}
  \SetKwInOut{Output}{Output}
%  \SetAlgoLined
  \Input{$\hat{\theta}_c$, $\hat{H}(\hat{\theta}_c)$,
    $\hat{J}(\hat{\theta}_c)$, $\theta_1 \in \Theta$, a proposal
    distribution $q(\cdot \mid \theta)$ and an adjusted composite
    likelihood $L_{\rm adj}(\cdot;y)$}
  \Output{A realisation of length $N+1$ from a Markov chain}
  \BlankLine
  \For{$t \leftarrow 1$ \KwTo $N$}{%
    $\theta^{(p)} \sim q(\cdot \mid \theta^{(t)})$\;
    $\alpha_{\rm adj}(\theta^{(t)}, \theta^{(p)})  \leftarrow \min\left\{1,
      {L_{\rm adj}(\theta^{(p)};y)\pi(\theta^{(p)}) q(\theta^{(t)}
        \mid \theta^{(p)})\over L_{\rm adj}(\theta^{(t)};y)\pi(\theta^{(t)})
        q(\theta^{(p)} \mid \theta^{(t)})} \right\}$\; 
    $U \sim U(0,1)$\;
    \eIf{$\alpha_{\rm adj}(\theta^{(t)}, \theta^{(p)}) \leq U$}{%
      $\theta^{(t+1)} \leftarrow \theta^{(p)}$\;
    }{$\theta^{(t+1)} \leftarrow \theta^{(t)}$\;}}
  \Return{$\{\theta^{(t)}\}_{t=1,\ldots,N+1}$}\;
  \caption{Adjusted Metropolis--Hastings algorithm.}
  \label{alg:adjMH}
\end{algorithm}

Implementation with one of the adjusted likelihoods, $ L_{\rm magn}(\theta;y)$ or $ L_{\rm curv}(\theta;y)$,  
requires only a preliminary maximisation of the composite likelihood
to estimate the matrices $H(\theta_0)$ and $J(\theta_0)$ for the adjustment. The
argument that establishes detailed balance for the original Metropolis--Hastings algorithm \citep[Theorem~7.2]{Robert2005}
applies to Algorithm~\ref{alg:adjMH}, and it can be shown that apart from normalizing
constants, the stationary distribution of the Markov chain is given by 
\begin{equation}
  \label{eq:statyMag}
  L^{\rm tot}_c(\theta;y)^k\pi(\theta),\quad k = p / \sum_{i=1}^p \hat{\lambda}_i,
\end{equation}
for the magnitude adjustment and 
\begin{equation}
  \label{eq:statyCurv}
  \exp \{\ell_{\rm curv}(\theta;y) \}\pi(\theta)
\end{equation}
for the curvature adjustment.  
The stationary distributions (\ref{eq:statyMag}) and (\ref{eq:statyCurv}) should provide better coverage than if an unadjusted composite likelihood was used. 

\subsection{Gibbs sampling}
\label{sec:adjust-gibbs-sampl}

When the unknown parameter $\theta$ has low dimension, Algorithm~\ref{alg:adjMH} should provide 
approximate inference for $\theta$ without too much Monte Carlo effort.  For models in
which $\theta$ is of high dimension, however, the probability of acceptance may be too low for Algorithm~\ref{alg:adjMH}  to be viable, and then the parameter vector is often partitioned and Gibbs sampler employed. Let us write $\theta =
(\theta_1^T, \ldots, \theta_G^T)^T$, where $\theta_j \in
\mathbb{R}^{p_j}$ and $\sum_{j = 1}^G p_j = p$, and suppose that we wish
to draw from 
\begin{equation}
\label{eqn:  Gibbs1}
\pi(\theta \mid y) \propto L(\theta ; y) \pi(\theta).
\end{equation}
A typical implementation of a Gibbs sampler will successively
draw from 
\begin{equation}
\label{eqn: Gibbs2}
\pi(\theta_j \mid \theta_{-j}, y) \propto L(\theta_j \mid \theta_{-j}, y) \pi(\theta_j),\quad j=1,\ldots, G, 
\end{equation}
where $\theta_{-j}$ is
the parameter vector $\theta$ with the elements of $\theta_j$ removed.  In this
section we propose two Gibbs samplers for use with
composite likelihoods.

\subsubsection{Overall Gibbs sampler}
\label{sec:overall-gibbs}

Since the true likelihood is unobtainable, we use the Gibbs sampler with 
an adjusted composite likelihood. 
We could replace $L(\theta ; y)$ in (\ref{eqn: Gibbs1})
with $L_{\rm adj}(\theta; y)$, where $L_{\rm adj}$ is either the 
magnitude- or the curvature-adjusted composite likelihood.  To perform 
Gibbs sampling, $\hat{\theta}_c$, $\hat{H}(\hat{\theta}_c)$ and 
$\hat{J}(\hat{\theta}_c)$ can be estimated once prior to running the 
algorithm, and $L_{\rm adj}(\theta; y)$ can be calculated.  Gibbs 
sampling then proceeds as usual.
%, with (\ref{eqn:  Gibbs2}) becoming
%$\pi(\theta_j \mid \theta_{-j}, y) \propto L_{\rm overall}(\theta_j ; y, 
%\theta_{-j}) \pi(\theta_j)$, where $L_{\rm overall}$ is simply the likelihood 
%associated with the marginal of $L_{\rm adj}(\theta; y)$.
%\footnote{Not sure quite what this means.}
%\footnote{Dan:  cut the end of this paragraph as I think it was confusing and not needed.  See Anthony's earlier comment}

As the Gibbs sampler is a special case of the
Metropolis--Hastings algorithm \citep[sec 10.2.2]{Robert2005}
and it was shown in Section~\ref{sec:adjust-metr-hast} that the latter could accommodate an adjusted
composite likelihood, this overall Gibbs sampler algorithm
converges to the stationary distributions given by 
\eqref{eq:statyMag} or \eqref{eq:statyCurv}.

%The Gibbs sampler is a special case of the
%Metropolis--Hastings algorithm \citep[sec 7.1.4]{Robert2005}, and this
%suggests an ``overall'' Gibbs sampler in which $\hat{\theta}_c$,
%$\hat{H}(\hat{\theta}_c)$ and $\hat{J}(\hat{\theta}_c)$ are estimated
%once prior to running the algorithm, and the sampler converges to the
%same stationary distribution, given by equations \eqref{eq:statyMag}
%or \eqref{eq:statyCurv}, as for the above Metropolis--Hastings
%algorithm.

\subsubsection{Adaptive Gibbs sampler}
\label{sec:adaptive-gibbs}

In real problems, the dimensions of $\theta$, and hence of $\hat
\theta_c$, can be quite large.  By finding $\hat{\theta}_c$,
$\hat{H}(\hat{\theta}_c)$ and $\hat{J}(\hat{\theta}_c)$ only once
before implementing the algorithm, the overall Gibbs sampler 
loses the `spirit' of Gibbs sampling, which is to sample the
lower-dimensional $\theta_j$ given the current value of $\theta_{-j}$.

%%In the next section, we detail a 
%%method which performs the adjustment to the likelihood of
%%$\theta_j$ given $\theta_{-j}$; that is, {\em within} the Gibbs sampler.
%Alternatively,  Metropolis--Hastings step can also be employed within a
%Gibbs sampler when the conditional distribution cannot be directly 
%sampled. This suggests a Gibbs implementation wherein the composite
%likelihood is adjusted at each step; the adjustments being conditional on 
%the current state $\theta_{-j}^{(t)}$ of the chain. We refer to this
%implementation, given in Algorithm~\ref{alg:adjGibbs}, as the
%``adaptive Gibbs sampler''.

An alternative to adjusting the likelihood in (\ref{eqn: Gibbs1}) is to replace the likelihood in
(\ref{eqn: Gibbs2}) by an 
adjusted composite likelihood.  That is, the likelihood for $\theta_{j}$ can be
adjusted based on the current values of $\theta_{-j}$.  Since this 
adjustment requires knowledge of the maximum composite likelihood estimates, the value of $\hat \theta_{j, c} \mid \theta_{-j} = \theta_{-j}^{(t)}$
 must be found at each step.  This approach has the  advantage that the adjusted
composite likelihood approximation using the current value
of $\theta_{-j}$ should be more accurate, as the approximation is made
in a lower-dimensional parameter space.
In particular if $\theta_j$ is scalar, then
the magnitude adjustment of the composite likelihood ratio statistic
is exact; see~(\ref{eq:asympAdjCompLikRJ}). This
adaptive Gibbs sampler is given in Algorithm~\ref{alg:adjGibbs}.
%\footnote{I'm a bit confused about this.  Algorithm 2 is called a Gibbs sampler, but it has an M-H rejection step.  Is this a copy-paste error from Algorithm 1, or is this algorithm mis-named?  Some further explanation needed?}

\begin{algorithm}
%  \LinesNumbered
  \SetKwInOut{Input}{Input}
  \SetKwInOut{Output}{Output}
%  \SetAlgoLined
  \Input{$\theta^{(1)} \in \Theta$}
  \Output{A realisation of length $N+1$ from a Markov chain}
  \BlankLine
  \For{$t \leftarrow 1$ \KwTo $N$}{%
    \For{$j \leftarrow 1$ \KwTo $G$}{%
      Get the restricted maximum composite likelihood estimate 
      $\hat{\theta}_{j,c}$ with $\theta_{-j}$ held fixed at $\theta_{-j}^{(t)}$\;
      Get $\hat{H}_{j,j}(\hat{\theta}_j) = \nabla^2 \ell_c(\hat \theta_{j,c} \mid \theta_{-j}^{(t)}, y)$
      and $\hat{J}_{j,j}(\hat{\theta}_j)$, the sample covariance matrix of $\nabla \ell_c(\hat \theta_{j,c} \mid  \theta_{-j}^{(t)}, y_i), i = 1, \ldots n$, and define the adjusted composite
      log-likelihood $\ell_{\rm adj}(\theta_j \mid \theta_{-j}^{(t)}, y)$ from either (\ref{eq:adjCompLikRJ})  or (\ref{eq:adjlclik});\\
      Draw $\theta_j^{(t+1)}$ from  $L_{\rm adj}(\theta_j ; y, \theta_{-j}^{(t)}) \pi(\theta_j \mid \theta_{-j})$ (using Metropolis--Hastings updates if necessary);
%      $\theta^{(p)}_j \sim q(\cdot \mid \theta^{(t)}_j)$\;
%      $\alpha_{\rm adj}(\theta^{(t)}_j, \theta^{(p)}_j) \leftarrow \min \left\{1,
%        \frac{L_{\rm adj}(\theta^{(p)}; y) \pi(\theta^{(p)}_j)
%          q(\theta^{(t)}_j \mid
%        \theta^{(p)}_j)}{L_{\rm adj}(\theta^{(t)}; y)
%          \pi(\theta^{(t)}_j) q(\theta^{(p)}_j \mid \theta^{(t)}_j)}
%      \right\}$\;
%      $U \sim U(0,1)$\;
%      \eIf{$\alpha_{\rm adj}(\theta^{(t)}_j, \theta^{(p)}_j) \leq U$}{%
%        $\theta^{(t+1)}_j \leftarrow \theta^{(p)}_j$\;
%      }{$\theta^{(t+1)}_j \leftarrow \theta^{(t)}_j$\;}

    }
  }
  \Return{$\{\theta^{(t)}\}_{t=1,\ldots,N+1}$}\;
  \caption{Adaptive adjusted Gibbs sampler.}
  \label{alg:adjGibbs}
\end{algorithm}

It can be shown that Algorithm~\ref{alg:adjGibbs} corresponds to a 
well-defined posterior by considering 
%To provide for justification for this procedure, consider 
the completion \citep[section~10.1.2]{Robert2005}:
\begin{equation*}
  \pi(\hat \theta, \theta \mid y) = \prod_{j = 1}^G \pi(\hat
  \theta_j \mid \theta, y) \pi(\theta \mid y),
\end{equation*}
where $\pi(\theta\mid y)$ represents the target density. 
Note that
\begin{equation*}
  \pi(\theta \mid y) = \int  \prod_{j = 1}^G  \pi(\hat \theta_j
  \mid \theta, y) \pi(\theta \mid y) d\hat \theta
\end{equation*}
as required for a completion, provided that $\pi(\hat \theta_j \mid
\theta, y)$ is a valid density.  Define
\begin{equation*}
  \pi(\hat \theta_j \mid \theta, y) = \delta_{\arg \max
  L_c(\theta_j \mid \theta_{-j}, y)}(\hat{\theta}_j),
\end{equation*}
that is, a Dirac measure on the value of $\theta_j$ that maximizes the
composite likelihood given the current values of $\theta_{-j}$.  
If the maximum composite likelihood estimates $\hat \theta_{j}$ could
be found analytically, then Algorithm~\ref{alg:adjGibbs} would simply 
be a Gibbs sampler on the completion.
%\footnote{Would it?  It looks to me as though $\pi(\theta\mid y)$ is changing at each step of the algorithm, because the adjusted composite likelihood is changing.}  
Since  $\hat \theta_{j}$ must 
be obtained numerically, convergence of the Markov chains must be
carefully checked by examining the output.
%\footnote{Does this mean that there is a theoretical check to be made, or that given output must be checked?}
%\footnote{Dan:  made change in response to Anthony's comment.}

%%The Gibbs sampler can then be implemented with the
%%completion; see Algorithm~\ref{alg:adjGibbs}. 
%This second Gibbs
%sampler has a different stationary distribution to the first Gibbs
%sampler/Metropolis--Hastings algorithm. An especially interesting
%feature is that the adaptive version of the magnitude adjustment is no
%longer an overall adjustment as said in
%Section~\ref{sec:prop-adjustm}, because each element of $\theta$ is
%adjusted differently.

In the context of the adaptive Gibbs sampler, both the magnitude
and curvature adjustments 
%in Sections~\ref{sec:magnitude-adjustment} and~\ref{sec:curvature-adjustment} 
must be understood as adjusting the conditional likelihood $L_c(\theta_j \mid \theta_{-j},
y)$.  That is, $k$ in equation (\ref{eq:adjCompLikRJ}) now becomes
$p_j / \sum_{i = 1}^{p_j} \hat \lambda_i$ where $\hat \lambda_i$ are
the eigenvalues of the matrix defined by $\hat H(\hat \theta_j)$ and
$\hat J(\hat \theta_j)$.  Similarly, the matrix $C$ in
\eqref{eq:mtxCDefn} is defined by $\hat H(\hat \theta_j)$ and $\hat
J(\hat \theta_j)$.

It is instructive to tie each of the Gibbs samplers to the asymptotic distribution of the posterior.
Let 
%$\pi_{\rm true}(\theta \mid y)$
$\pi(\theta \mid y)$
denote the composite posterior distribution evaluated at $\theta \in \mathbb{R}^p$ and further assume that the asymptotic posterior distribution corresponds with that of the maximum composite likelihood estimator,
%\footnote{What does the proportional with a dot symbol mean?  Dan:  we need a symbol for "asymptotically proportional to"}
%\begin{equation}
%\label{eqn:  truePost}
%  \log \pi_{\rm true}(\theta \mid y) \; \dot \propto \; -\frac{1}{2} (\theta - \hat \theta_{\rm true})^T H^{\rm true}(\hat \theta_{\rm true}) (\theta - \hat \theta_{\rm true}).
%\end{equation}
\begin{equation}
\label{eqn:  truePost}
 % \log \pi_{\rm true}(\theta \mid y) 
    \log \pi(\theta \mid y) 
  \; \dot \propto \; -\frac{1}{2} (\theta - \theta_0)^T H(\theta_0) J^{-1}(\theta_0) H(\theta_0) (\theta - \theta_0),
\end{equation}
where $\dot \propto $ means `asymptotically proportional to'. 
%where 
%$\hat \theta_{\rm true}$ is the composite posterior mode and 
%$H(\theta)= - \nabla^2 \log \pi_{\rm true}(\theta \mid y)$
%and $J(\theta)= - \nabla^2 \log \pi_{\rm true}(\theta \mid y)$.
Gibbs sampling for a given partition $\theta = (\theta_j, \theta_{-j})^T$, where $\theta \in \mathbb{R}^{p_j}$ and $\theta_{-j} \in \mathbb{R}^{p - p_j}$, would involve drawing from 
%$\pi_{\rm true}(\theta_j \mid \theta_{-j}, y)$, 
$\pi(\theta_j \mid \theta_{-j}, y)$, 
 the conditional posterior distribution of $\theta_j$ given some fixed value for $\theta_{-j}$.

In the overall Gibbs sampler, one begins by approximating (\ref{eqn:  truePost}) with
\begin{equation}  
\label{eqn:  overallPost}
   \log \pi_{\rm adj}(\theta \mid y) \; \dot \propto \; -\frac{1}{2} (\theta - \hat \theta_c)^T H^{\rm adj}(\hat \theta_c) (\theta - \hat \theta_c).
\end{equation}
where $H^{\rm adj}(\hat \theta_c)^{-1}$ is the covariance matrix in (\ref{eq:asymptMagnPostDist}) or (\ref{eq:asymptCurvPostDist}) for the magnitude- and curvature-adjusted posteriors respectively.

Let $\hat \theta_c = (\hat \theta_{c,j}, \hat \theta_{c,-j})^T$ and partition
$$
  H^{\rm adj}(\hat \theta_c) = 
  \left[
    \begin{array}{c c}
      H^{\rm adj}_{j,j}(\hat \theta_c) & H^{\rm adj}_{j,-j}(\hat \theta_c)\\
      H^{\rm adj}_{-j,j}(\hat \theta_c) & H^{\rm adj}_{-j,-j}(\hat \theta_c)
    \end{array}
  \right].
$$
%That is, $\hat \theta_j$ corresponds to the first $j$ elements of $\hat \theta$.

Since (\ref{eqn: overallPost}) implies  that $\pi_{\rm adj}\{ (\theta_{j}, \theta_{-j})^T \}$ is approximately a Gaussian density with mean $(\hat \theta_{c,j}, \hat \theta_{c,-j})^T$ and covariance matrix $\Sigma = H^{\rm adj}(\hat \theta_c)^{-1}$, we see that 
$\pi_{\rm adj}(\theta_j \mid \theta_{-j}, y)$ is approximately Gaussian, with mean 
\begin{equation}
\label{eqn:  overallMean}
  \hat \theta_{c,j} + \Sigma_{j,-j} \Sigma_{-j, -j}^{-1} (\theta_{-j} - \hat \theta_{c,-j}) =  \hat \theta_{c,j} - H^{\rm adj}_{j,j}(\hat \theta_c)^{-1} H^{\rm adj}_{j,-j}(\hat \theta_c) (\theta_{-j} - \hat \theta_{c,-j})
\end{equation}
and covariance matrix
\begin{equation}
\label{eqn:  overallVar}
  \Sigma_{j,j} - \Sigma_{j,-j} \Sigma_{-j, -j}^{-1} \Sigma_{-j, j} = H^{\rm adj}_{j,j}(\hat \theta_c)^{-1}.
\end{equation}

The adaptive Gibbs sampler makes its approximation later in the algorithm. 
Starting from (\ref{eqn: truePost}),
let $\theta_{-j}$ be given and consider $\log \pi(\theta_{j} \mid  \theta_{-j}, y)$.
By partitioning $\theta_0$ and $H(\theta_0) J^{-1}(\theta_0) H(\theta_0) $, 
it is straightforward to show that the asymptotic conditional posterior is
\begin{equation}
\label{eqn: condtlPostDist}
   \log \pi(\theta_{j} \mid  \theta_{-j}, y) \; \dot \propto \; 
   (\theta_j - \mu_{j \mid -j})^T \Sigma_{j \mid -j}^{-1} (\theta_j - \mu_{j \mid -j}), 
\end{equation}
where
\begin{equation}
\label{eqn: condtlPostMean}  
 \mu_{j \mid -j} = \theta_{0,j} - \left\{ H(\theta_0) J^{-1}(\theta_0)
   H(\theta_0) \right\}_{j,j}^{-1}\left\{ H(\theta_0) J^{-1}(\theta_0)
   H(\theta_0) \right\}_{j,-j} (\theta_{-j} - \theta_{0, -j}),
\end{equation}
and 
\begin{equation}
\label{eqn: condtlPostVar}
 \Sigma_{j \mid -j} = \left\{ H(\theta_0) J^{-1}(\theta_0) H(\theta_0)
 \right\}_{j,j}^{-1}
\end{equation}
analogous to (\ref{eqn: overallMean}) and ({\ref{eqn: overallVar})
  above.  The adaptive Gibbs sampler makes its approximation to the
  conditional distribution, estimating the conditional mean by finding
  $\hat \theta_{c, j \mid -j}$, the value which maximizes the, lower-dimensional, conditional composite log-likelihood
  $ \ell_c(\theta_{j,c} \mid \theta_{-j}^{(t)}, y)$, and then
  adjusting this lower-dimensional likelihood to obtain an estimate for
  $\left\{ H(\theta_0) J^{-1}(\theta_0) H(\theta_0) \right\}_{j,j}$.
%$H^{\rm true}_{j,j}(\hat \theta_{\rm true})^{-1}$.

The advantage of the overall Gibbs sampler is computational and in its
simplicity; the adaptive Gibbs sampler's need to estimate $\hat
\theta_j$ at every step slows it tremendously.  The potential gain from the latter is that the approximation made by employing a
composite likelihood is made only for the subvector 
$\theta_j$ and is done with knowledge of the current values of the
other parameters.  In the next section we explore by simulation whether the adaptive Gibbs sampler improves overall
estimation.

\section{Simulation study}
\label{sec:simulation-study}

In this section, we use simulation to assess the performance of the
magnitude and the curvature adjustments. Following
\citet{Monahan1992}, we assess whether our adjustments yield
posteriors that are valid by coverage, i.e.,\ whether $\Pr[\theta \in
\mbox{CI}_\alpha(Y)] = \alpha$, under some probability measure for
$\theta$ defined on $\Theta$ and some credible intervals
$\mbox{CI}_\alpha$ with level $0 \leq \alpha \leq 1$.

We first apply the proposed adjustments to the stationary isotropic
Gaussian process of Section~\ref{sect:1.3} and compare the results obtained
using the adjusted composite likelihood to those using both the full
likelihood and the naive composite likelihood. We then focus
on spatial extremes by considering a Bayesian hierarchical model
involving max-stable processes.%  Some theoretical
% details for the estimation of the matrices $\hat H(\hat \theta_c)$ and
% $\hat J(\hat \theta_c)$ are given in
% Appendix~\ref{sec:estim-hthet-jthet}.

\subsection{Gaussian processes}
\label{sec:gaussian-processes}

We again consider a one-dimensional stationary Gaussian process with
mean $\mu \in \mathbb{R}$ and an exponential covariance function
$\gamma(h) = \tau \exp(-h/\omega)$, $\tau >0$, $\omega > 0$. We
examine two different forms of dependence, allowing $\omega$ to equal
$3$ and $1.5$, which respectively yield effective ranges for the
covariance of roughly $9$ and $4.5$.  The priors on $\mu$, $\tau$ are
those reported in Section~\ref{sec:introduction} while an inverse
Gamma density with shape $1/10$ and scale $1$ is assumed on $\omega$.
% \footnote{But there $\omega$ was fixed so had no prior.}
The stochastic process is replicated $n=50$ times in each simulation
and is observed at $K=20$ locations uniformly generated in the
interval $[0,20]$. The simulation was repeated $500$ times to assess
coverage, with $\mu = 0$ and $\tau = 1$ in each case.

\begin{figure}
  \centering
  \includegraphics[angle=-90,width=\textwidth]{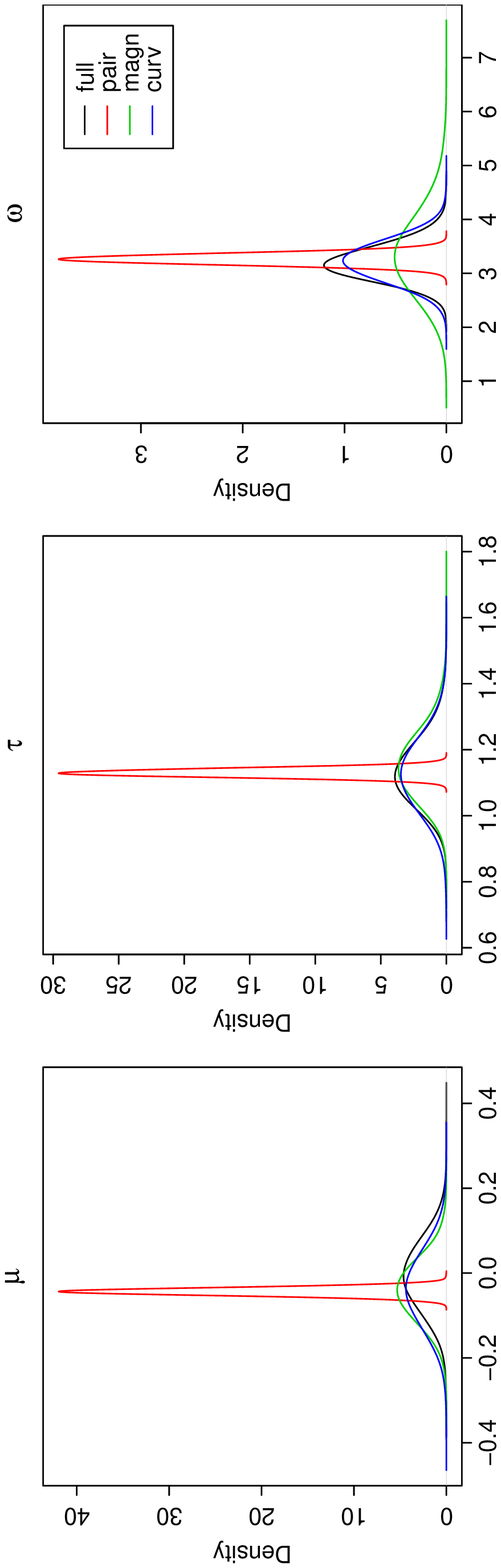}
  \includegraphics[angle=-90,width=\textwidth]{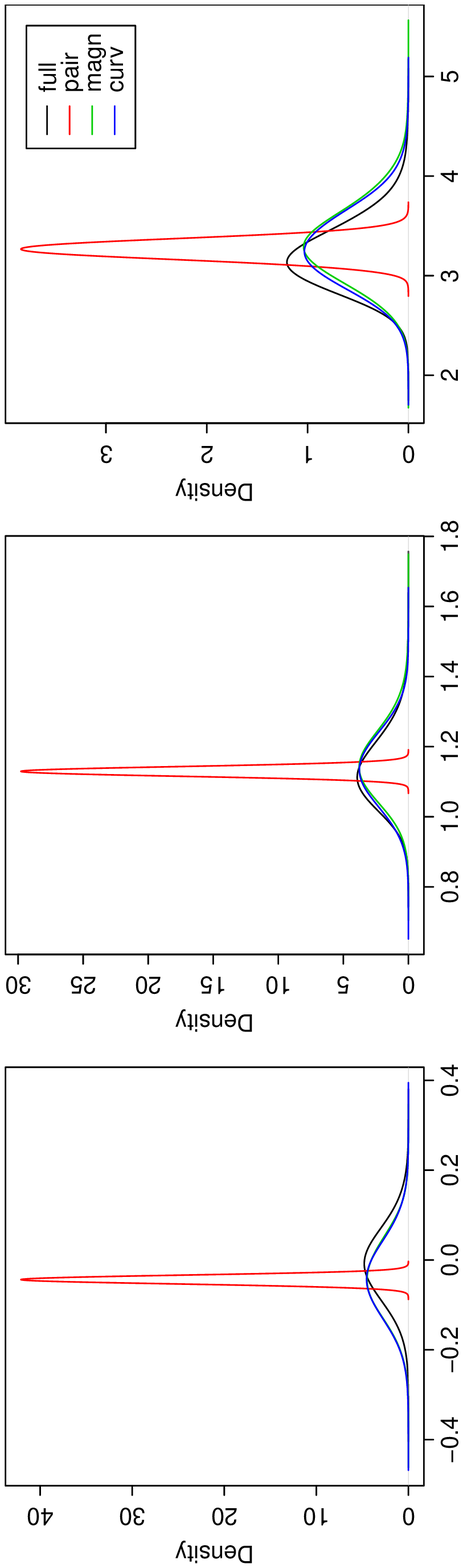}
  \caption{Comparison between the marginal full posterior (black), the
    marginal pairwise posterior (red) and the marginal adjusted
    pairwise posterior densities based on the magnitude (green) and
    curvature (blue) adjustments. The posterior distributions are
    derived from $n=50$ realisations of a Gaussian process having an exponential covariance function with
    $\mu=0$, $\tau = 1$ and $\omega = 3$ and  observed at
    $K=20$ locations. Top row:
    Metropolis--Hastings algorithm. Bottom row: Adaptive adjusted
    Gibbs sampler.}
  \label{fig:RJCBapproaches}
\end{figure}

Figure~\ref{fig:RJCBapproaches} compares the posterior densities
obtained from the full likelihood, the unadjusted pairwise posterior,
and the adjusted composite posterior distributions using the magnitude
and the curvature adjustments from a single simulation. There is a
large improvement due to the adjustment.  Owing to the
asymptotic unbiasedness of the maximum composite likelihood estimator,
the modes of the marginal composite posterior distributions are close
to those obtained from the full likelihood. The use of the adaptive
Gibbs sampler for the magnitude adjustment seems to improve the
approximation to the full posterior, particularly for the range
parameter $\omega$; recall from Section~\ref{sec:adjust-gibbs-sampl}
that this is not an overall magnitude adjustment.  The adaptive sampler used here has three blocks each comprising a single parameter.

\begin{table}
  \centering
  \caption{Empirical coverages (\%) for nominal $95\%$ credible intervals based on
    $500$ Gaussian process simulations.  ``Full'' denotes coverage
    with the full posterior, ``Magnitude'' corresponds to the magnitude
    adjusted posterior, ``Curvature'' to the curvature adjusted posterior,
    and ``Unadjusted'' to the naive composite posterior.}
  \label{tab:covPerf}
  \begin{tabular}{lccccccccccccccc}
    \hline
    \multicolumn{16}{c}{Metropolis--Hastings}\\
    & \multicolumn{3}{c}{Full} && \multicolumn{3}{c}{Magnitude} &&
    \multicolumn{3}{c}{Curvature} && \multicolumn{3}{c}{Unadjusted}\\
    \cline{2-4} \cline{6-8} \cline{10-12} \cline{14-16}
    & $\mu$ & $\tau$& $\omega$ && $\mu$ & $\tau$& $\omega$ && $\mu$ &
    $\tau$& $\omega$ && $\mu$ & $\tau$& $\omega$\\ 
    \hline
    %Dan's sim
    $\omega = 3$ & 96 &  94 & 94 && 89 & 92 & 100 &&
    94 &  93 & 94 && 16 & 21 & 37 \\
    $\omega = 1.5$ & 94 & 95 & 96 && 85 & 93 & 100 &&
    94 & 94 & 93 && 19 & 22 & 53\\ 
    \hline
    \hline
    \multicolumn{16}{c}{Overall Gibbs sampler}\\
    & \multicolumn{3}{c}{Full} && \multicolumn{3}{c}{Magnitude} &&
    \multicolumn{3}{c}{Curvature} && \multicolumn{3}{c}{Unadjusted}\\
    \cline{2-4} \cline{6-8} \cline{10-12} \cline{14-16}
    & $\mu$ & $\tau$& $\omega$ && $\mu$ & $\tau$& $\omega$ && $\mu$ &
    $\tau$& $\omega$ && $\mu$ & $\tau$& $\omega$\\ 
    \hline
    $\omega = 3$ & $95$ & $96$ & $95$ &&  $87$ & $93$ &
    $100$ && $94$ & $94$ & $90$ && $19$ & $16$ & $41$\\
    $\omega = 1.5$ & $96$ & $96$ & $96$ &&  $87$ & $94$ &
    $100$ && $94$ & $94$ & $94$ && $23$ & $21$ & $55$\\     
    \hline
    \hline
    \multicolumn{16}{c}{Adaptive Gibbs sampler}\\
    & \multicolumn{3}{c}{Full} && \multicolumn{3}{c}{Magnitude} &&
    \multicolumn{3}{c}{Curvature} && \multicolumn{3}{c}{Unadjusted}\\
    \cline{2-4} \cline{6-8} \cline{10-12} \cline{14-16}
    & $\mu$ & $\tau$& $\omega$ && $\mu$ & $\tau$& $\omega$ && $\mu$ &
    $\tau$& $\omega$ && $\mu$ & $\tau$& $\omega$\\ 
    \hline
    $\omega = 3$ & $96$ & $94$ & $95$ &&  $95$ & $92$ &
    $93$ && $95$ & $94$ & $93$ && $20$ & $24$ & $39$\\
    $\omega = 1.5$ & $95$ & $96$ & $95$ &&  $95$ & $95$ &
    $95$ && $94$ & $97$ & $95$ && $17$ & $24$ & $55$\\     
    \hline
  \end{tabular}
\end{table}

Table~\ref{tab:covPerf} summarizes the empirical coverages based on
$500$ replicate data sets.  Overall, the adjustments give reasonable
credible intervals, whereas the naive composite posterior has poor
coverage. The Metropolis--Hastings algorithm and overall Gibbs sampler have the same stationary
distribution and give the
same coverages for each adjustment. The curvature adjustment performs better overall than
the magnitude adjustment, particularly for the mean and range
parameters $\mu$ and $\omega$. The improvement in coverage due to using the adaptive Gibbs sampler 
appears greater for the magnitude adjustment than for the
curvature adjustment, partly because there is more room for improvement, and because the latter was already adjusting each element of
$\theta$ differently. The curvature and adaptive magnitude adjustments
yield the best coverages.

\begin{figure}
  \centering
  \includegraphics[angle=-90,width=\textwidth]{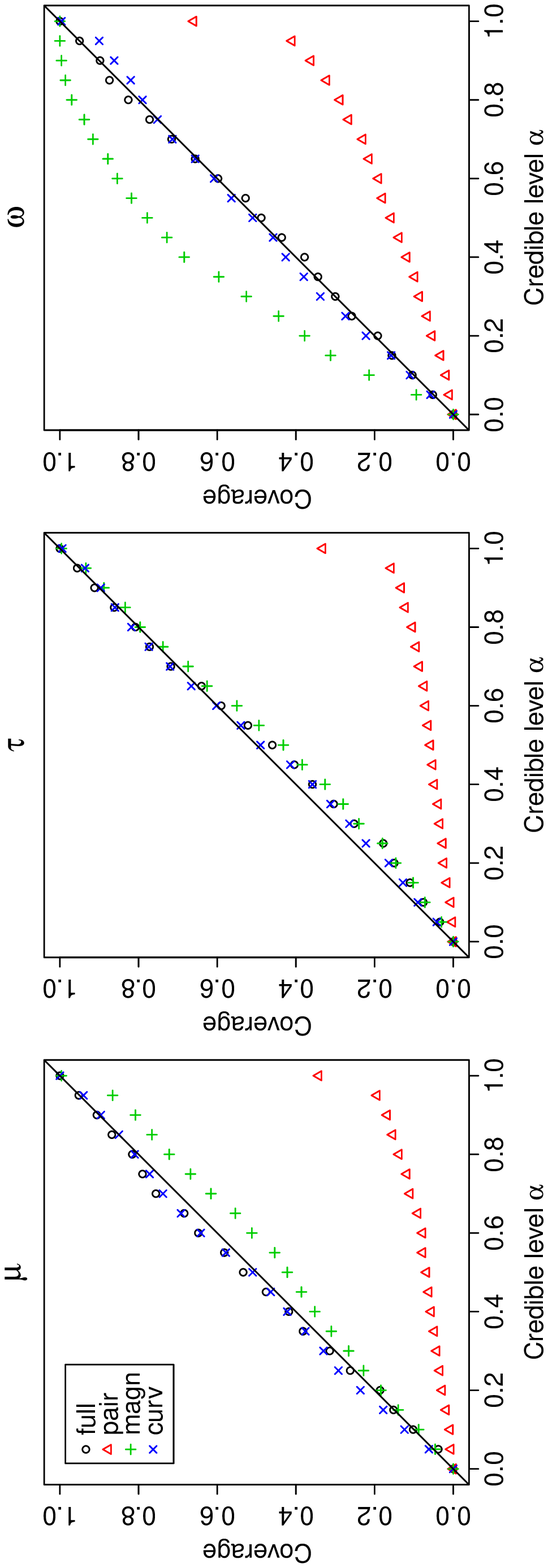}
  \includegraphics[angle=-90,width=\textwidth]{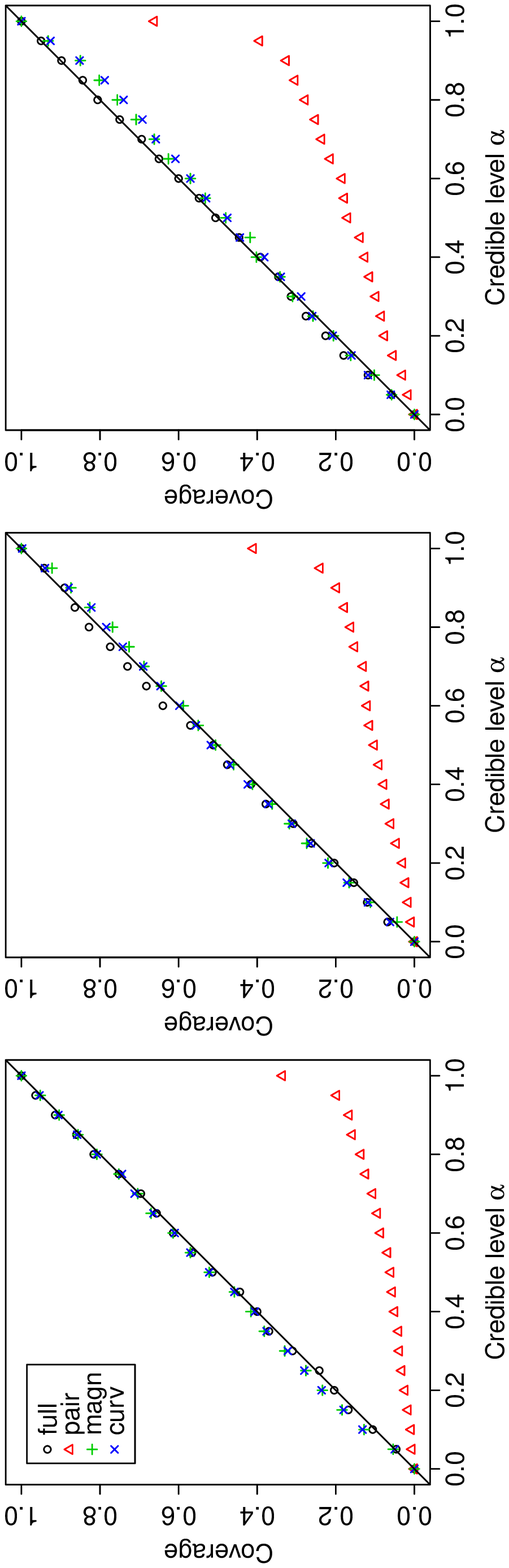}
  \caption{Variation of the empirical coverages with the credible
    level $\alpha$, based on $500$ replicates of the Gaussian process
    simulation with $\mu = 0$, $\tau = 1$ and $\omega = 3$, for the
    full, the non adjusted pairwise and the magnitude/curvature
    adjusted posteriors. Top row: Overall Gibbs sampler. Bottom row:
    Adaptive Gibbs sampler.}
  \label{fig:coverage}
\end{figure}

Figure~\ref{fig:coverage}, which complements Table~\ref{tab:covPerf}
by showing how the empirical coverages depend on the credible level
for the overall and adaptive Gibbs samplers, corroborates the
conclusions drawn from Table~\ref{tab:covPerf}. Compared to the
unadjusted composite posterior, the proposed adjustments clearly
improve the coverages and seem to yield essentially the same coverages as
the full posterior, though the latter provides shorter intervals, if it is available. The adaptive Gibbs sampler for the magnitude
adjustment performs better than its overall counterpart, indicating
that the latter might not be flexible enough to provide the correct
coverages for each element of the parameter vector. The curvature
adjustment again seems to be improved less by the adaptive version of
the Gibbs sampler.%\footnote{Mathieu: They were 3 blocks of size one (Anthony: mentioned above) and for the new  simulation only 2 $\{\mu\}$ and $\{\tau, \omega\}$. We may want to  rewrite a little bit this part.  Anthony: is the new simulation mentioned here in the text?}

\begin{figure}
\begin{center}
  \includegraphics[angle=-90,width=\textwidth]{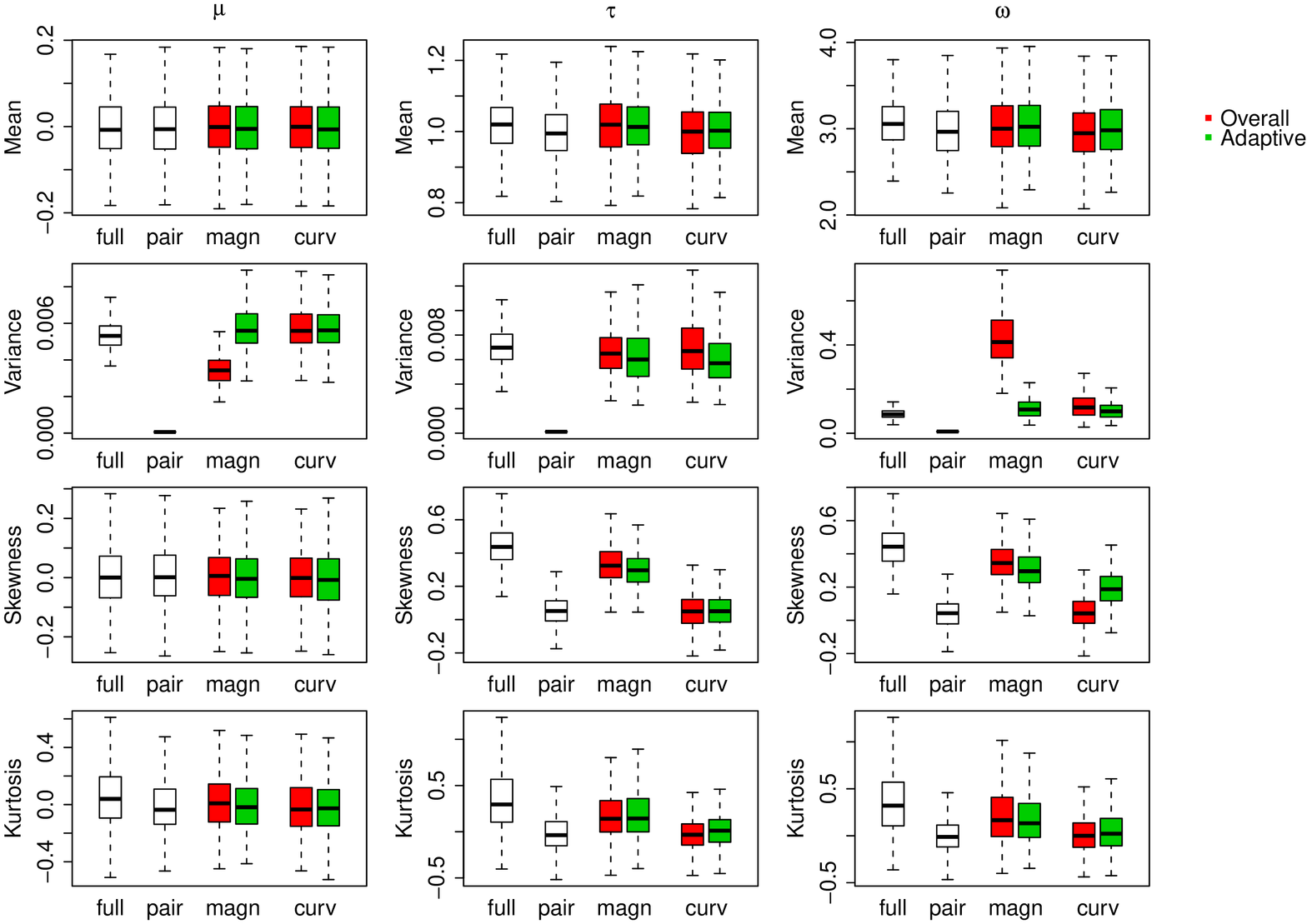}
  \caption{Boxplots of the sample centered moments of the estimated
    posterior distribution for each of the $500$ simulations ($\mu =
    0$, $\tau = 1$, $\omega = 3$) for the full posterior (full), the
    unadjusted pairwise posterior (pair), the magnitude adjusted
    composite posterior (magn) and the curvature adjusted composite
    posterior (curv). Red boxplots: Overall Gibbs sampler. Green
    boxplots: Adaptive Gibbs sampler.}
\label{fig:momentPlot}
\end{center}
\end{figure}

Figure~\ref{fig:coverage} shows that the proposed adjustments have
good coverage properties, but it is also interesting to check to what
extent the composite posterior distributions share common features
with the full posterior. Figure~\ref{fig:momentPlot} shows boxplots of
the first four centered moments of the estimated posterior
distributions. As one would expect from the fact that the composite
likelihoods give unbiased estimating equations, the first moments of
the composite posterior distributions, including the unadjusted one,
match those of the full posterior. The
variance of the unadjusted pairwise posterior
distribution is much too small, but those of the adjusted posterior distributions
are closer to that of the full posterior. The
magnitude adjustment combined with the overall Gibbs sampler has a
smaller variance for the mean $\mu$ and a larger one for the range
$\omega$; this clarifies why Table~\ref{tab:covPerf} shows that this
particular adjustment tends to undercover $\mu$ and overcover
$\omega$.  Except for $\mu$, none of the adjustments gives the
correct skewness and kurtosis, though the magnitude adjustment is
slightly better.  Nevertheless, both adjustments can capture the first
two moments well, and despite the degradation of the approximation
with distance from $\theta_0$, yield coverage rates which are very
comparable to those obtained using the full likelihood.

Finally, we investigate the difference between the magnitude- and 
curvature-adjusted posteriors and the effect of the dimension of the blocks 
used in the adaptive Gibbs sampler.  
As noted in Section \ref{sec:magnitude-adjustment}, the magnitude 
adjustment will recover the $\chi^2$ null
distribution only if the dimension of 
$\theta_j$ is one.
In addition to running the adaptive Gibbs sampler with $\mu, \tau,$ and 
$\omega$ each serving as its own block, we
also ran a two-block version of the adaptive Gibbs sampler with 
$\theta_1 = \mu$ and $\theta_2 = (\tau, \omega)^T$.
For the individual block version of the adaptive Gibbs sampler, 
there is virtually no difference in the estimates of the magnitude-
and curvature-adjusted posteriors, reflecting that both adjustments
adequately capture the information contained in the composite likelihood.
However, for the two-block version of the sampler, the empirical posterior correlations of $\tau$ and $\omega$ differ: the curvature-adjusted posterior 
gives ${\rm Cov}(\tau,\omega)\approx 0.69$,  whereas the magnitude-adjusted posterior gives ${\rm Cov}(\tau,\omega)\approx 0.33$.
It is difficult to estimate both the sill and range parameters
of a Gaussian process \citep{ZhangH04}, whose ratio 
$\tau/\omega$ is important for applications such as interpolation.
The 95\% credible intervals for this ratio have an empirical
coverage rate of 96\% for the curvature-adjusted posterior, but a coverage
rate of 100\% for the magnitude-adjusted posterior.
This suggests that for the two-block
Gibbs sampler, the magnitude adjustment fails to fully capture the relationship
between these two parameters, thus giving further evidence that the
curvature adjustment is to be preferred, since it seems to provide
output that can be used more
flexibly.%\footnote{Dan: please see tex code which includes the
         %reference for Zhang.}

%@article{ZhangH04,
%author = "H. Zhang",
%title = "Inconsistent Estimation and Asymptotically Equal Interpolations in Model-Based Geostatistics",
%journal = "Journal of the American Statistical Association",
%volume = "99",
%year = "2004",
%pages = "250-261",
%}

\subsection{Bayesian hierarchical model for spatial extremes}
\label{sec:bayes-hier-model-1}

Let $Y_m(x)$, $x \in \mathcal{D}$, $m\geq 1$ be independent
replications of a stochastic process.  Asymptotic theory for extremes
implies that, provided the limit exists and is non-degenerate, the
process
\begin{equation*}
\max_{m= 1, \ldots   n}a_n(x)^{-1}\{ Y_m(x) - b_n(x)\}
\end{equation*}
converges weakly to a max-stable process $Z(x)$ as $n \to +\infty$
\citep{deHaan1984}.  Given observations that arise as block (e.g., 
annual) maxima, it is therefore natural to approximate their joint
distribution using such a process.  The univariate marginal
distributions for such a process will be generalised extreme-value
(GEV) distributions, which depend on three parameters.

Although the general methodology we propose could be applied with any
max-stable model \citep{Smith1990,Schlather2002,Kabluchko2009}, 
we focus here on the Gaussian extreme value process of \citet{Smith1990},
\begin{equation}
  \label{eq:smithModel}
  Z(x) = \max_{k \geq 1} \zeta_k \varphi(x - s_k)
\end{equation}
where $\{(\zeta_k, s_k)\}_{k\geq1}$ are the points of a Poisson
process on $(0,\infty) \times \mathcal{D}$, with $\mathcal{D} \subset
\mathbb{R}^d$, having intensity $\mbox{d$\Lambda$}(\zeta, s) = \zeta^{-2}
\mbox{d$\zeta$ds}$, and $\varphi$ is the zero mean $d$-variate normal
density with covariance matrix $\Sigma$. As formulated, $Z(x)$ has
unit Fr\'echet margins and its bivariate and trivariate marginal distributions
can be used to construct a composite likelihood 
\citep{Padoan2010,Genton.Ma.Sang:2011}.

A simple approach
to fitting max-stable models is to employ a pairwise likelihood
\citep{Padoan2010,Gholamrezaee2010}. To account for non-stationarity
in the marginal distributions, it is convenient to assume that the GEV
parameters follow response surfaces that depend on location and on
covariates such as altitude.  Often, however, the available covariates
do not fully explain the variation of the marginal distribution over
the study region. One approach to capturing the regional effects is to
construct a hierarchical model in which the marginal parameters of the
extreme value distribution follow a stochastic process, such as a
Gaussian process, over the study region.

Our approach is to use a max-stable process model within a
hierarchical framework; the max-stable model provides a theoretically
justified model for the local dependence, i.e.,\ the spatial dependence
of the extremes, and the hierarchy allows for flexibility in modeling how
the regional effects influence the marginal behavior.  The difficulty is that the full likelihood is unavailable, and so
fully Bayesian inference cannot be performed.  Instead we employ one
of the adjusted MCMC samplers suggested in
Section~\ref{sec:mcmc-samplers}.

Our chosen model has the data-process-prior framework of
most hierarchical models:
\begin{eqnarray*}
  Z \mid \boldsymbol{\mu}, \boldsymbol{\sigma}, \boldsymbol{\xi},
  \Sigma~&\sim& \mbox{Smith's max-stable model},\\
  \boldsymbol{\mu} \mid \boldsymbol{\beta}_\mu, \tau_\mu, \omega_\mu
  &\sim& \mbox{GP}\left(\mathbf{X}_\mu \boldsymbol{\beta}_\mu,
    \gamma_\mu\right),\\
  \log \boldsymbol{\sigma} \mid \boldsymbol{\beta}_\sigma,
  \tau_\sigma, \omega_\sigma &\sim& \mbox{GP}\left(\mathbf{X}_\sigma
    \boldsymbol{\beta}_\sigma, \gamma_\sigma\right)\\
  \boldsymbol{\xi} \mid \boldsymbol{\beta}_\xi, \tau_\xi,
  \omega_\xi~&\sim& \mbox{GP}\left(\mathbf{X}_\xi
    \boldsymbol{\beta}_\xi, \gamma_\xi\right),
\end{eqnarray*}
where $\boldsymbol{\mu}$, $\boldsymbol{\sigma}$, $\boldsymbol{\xi}$
represent the three GEV parameters, $\mbox{GP}(m, \gamma)$ denotes a
Gaussian process with mean $m$ and covariance function $\gamma$,
the $\gamma_\cdot$'s represent exponential covariance functions with corresponding sill and range  parameters
$\tau_\cdot$ and $\omega_\cdot$, and the 
$\boldsymbol{\beta}_\cdot$ are regression coefficients associated to
the design matrices $\mathbf{X}_\cdot$.

The prior level places independent priors on all parameters introduced
at the process level. We take conjugate normal priors for all
regression parameters $\beta_\cdot$, conjugate inverse gamma priors
for the $\tau_\cdot$, gamma priors for the range
parameters $\omega_\cdot$ and a Wishart prior for the
covariance matrix $\Sigma$ appearing in the Smith model.  In all
cases, the prior variance is set to be large so that the prior
densities, though proper, are relatively flat.

We performed a simulation study to evaluate our
approach.  Gaussian processes were simulated for $\mu(x)$,
$\sigma(x)$, and $\xi(x)$, with $\mu(x)$ and $\sigma(x)$ dependent,
and with values similar to those found for annual maximum rainfall
data.  Then, 50 max-stable processes with marginals given by $\mu(x)$,
$\sigma(x)$, and $\xi(x)$ were simulated according to the Smith model.
Fifty locations were chosen and the 50 observations at each location
were used to fit four models:
\begin{description}
\item[M1] the hierarchical model with a conditional independence
  assumption in the data layer, yielding a
  product of $K$ independent GEV densities, analogous to
  \citet{Cooley2007} or \citet{Sang2009};
\item[M2] the max-stable process hierarchical model with no
  adjustment;
\item[M3] the max-stable process hierarchical model with an adaptive
  curvature-adjusted Gibbs sampler; and
\item[M4] the max-stable process model where the marginals are
  described by a response surface in the covariates $x$, as proposed by
  \citet{Padoan2010}.
\end{description}

\begin{figure}
  \centering
  \includegraphics[angle=-90,width=\textwidth]{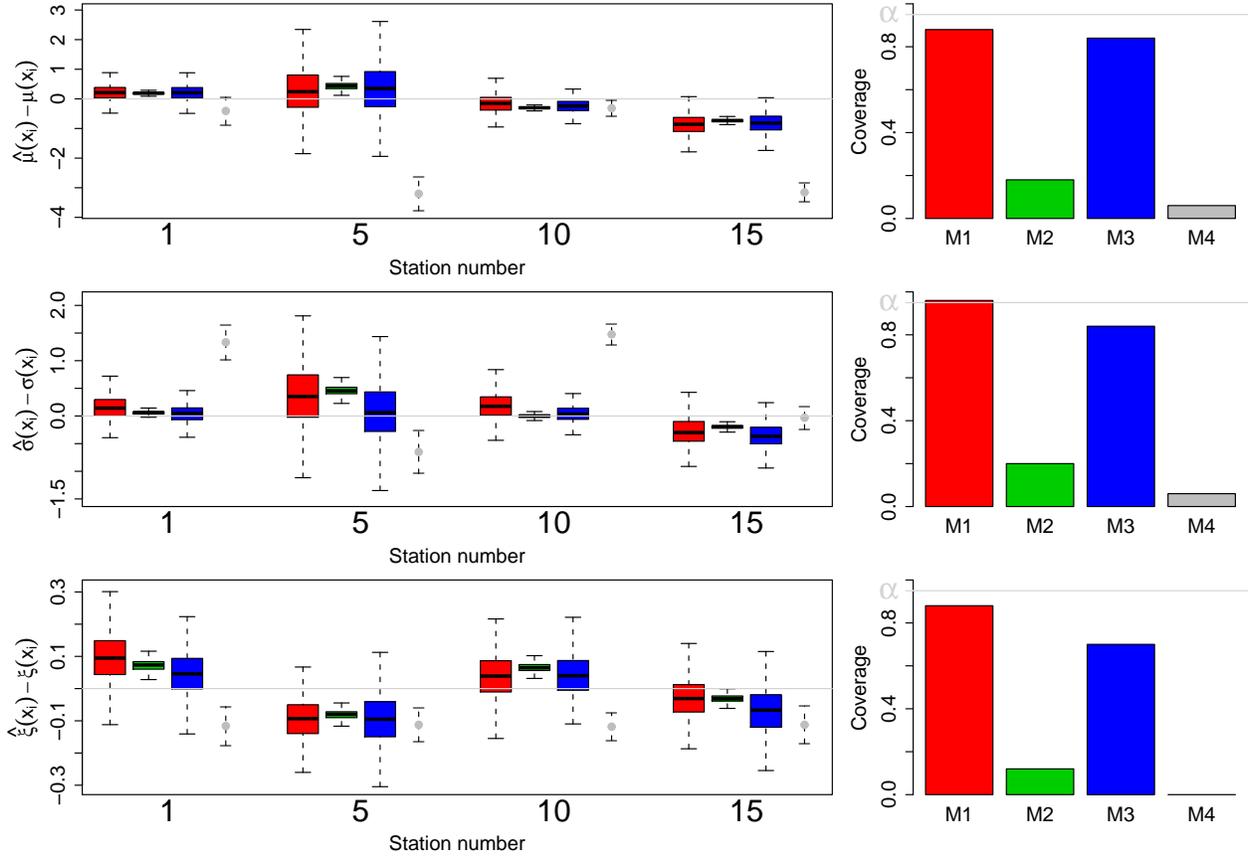}
  \caption{Boxplots of the difference between the true GEV parameters
    and all the states of the Markov chains for four stations (left
    panel). For each of the stations the boxplots are (from left to
    right) the conditional independence model (M1, red), the
    non-adjusted hierarchical model (M2, green), the hierarchical
    model with the curvature adjustment within an adaptive Gibbs
    sampler (M3, blue), and the asymptotic 95\% confidence limit from
    the max-stable response surface model (M4, grey).  The right panel
    shows the proportion of the credible intervals at level
    $\alpha=95\%$ containing the true GEV parameters.}
  \label{fig:CheckMarginsSim}
\end{figure}

The left panels of Figure~\ref{fig:CheckMarginsSim} show boxplots of
the differences between the true GEV parameters and all the states of
the Markov chains for four different stations, with the asymptotic
$95\%$ confidence intervals for the max-stable response surface model.
The right panels of Figure~\ref{fig:CheckMarginsSim} display the
coverage rates, for all 50 stations, of the $95\%$ posterior credible
intervals for the three hierarchical models, with the $95\%$
confidence intervals for the max-stable response surface model. As expected, the
unadjusted max-stable hierarchical model produces a posterior that is
too concentrated and yields very poor coverages, and the max-stable
trend surface model is not flexible enough to account for the
complicated regional behavior of the GEV parameters, as evidenced by
the poor point estimates in the box plots and the corresponding poor
coverage rates. The adjusted max-stable hierarchical model and the
conditionally independent hierarchical model produce very similar
posterior distributions and have similar coverage rates, although the
max-stable model does slightly less well.

\begin{figure}
  \centering
  \includegraphics[angle=-90,width=.32\textwidth]{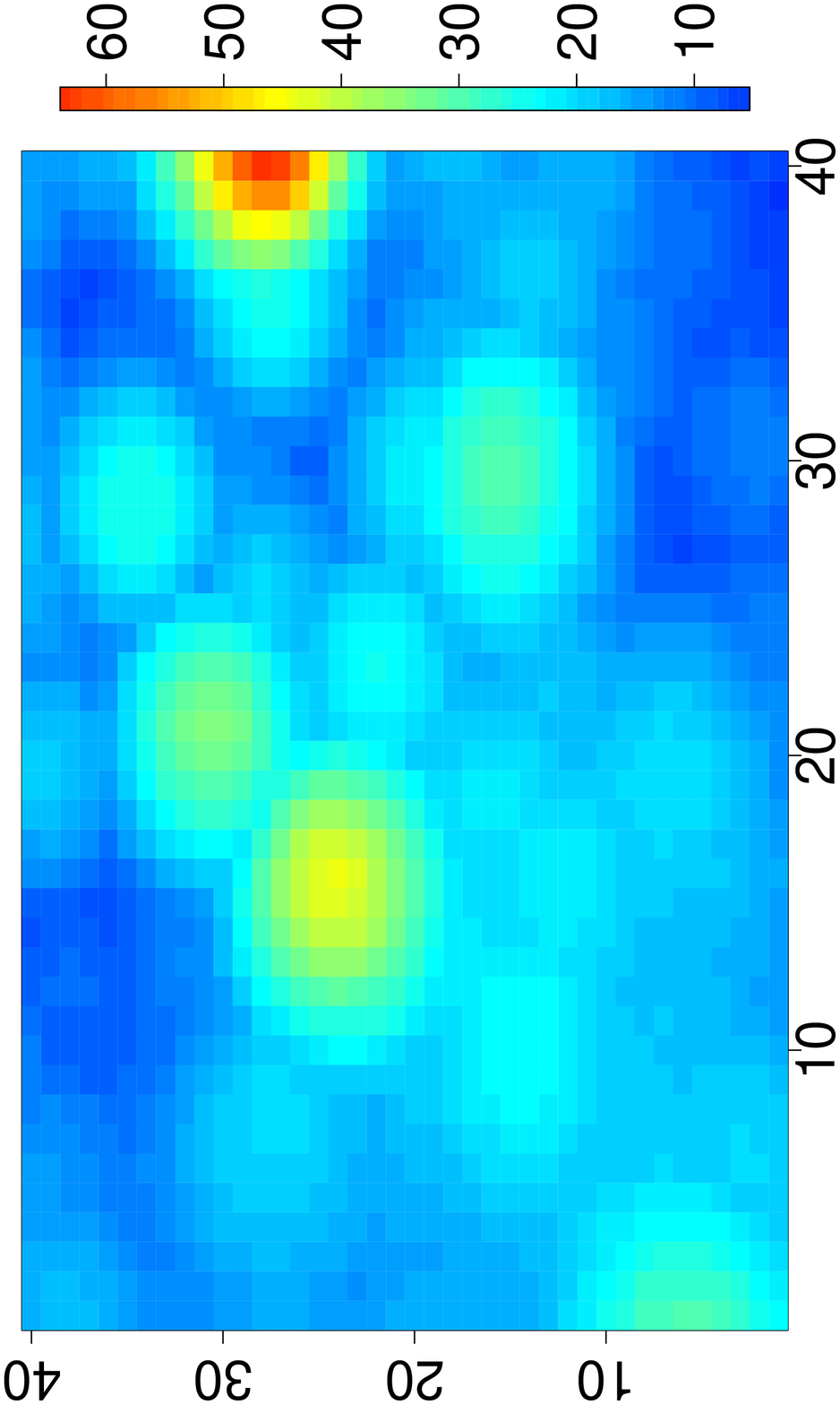}
  \includegraphics[angle=-90,width=.32\textwidth]{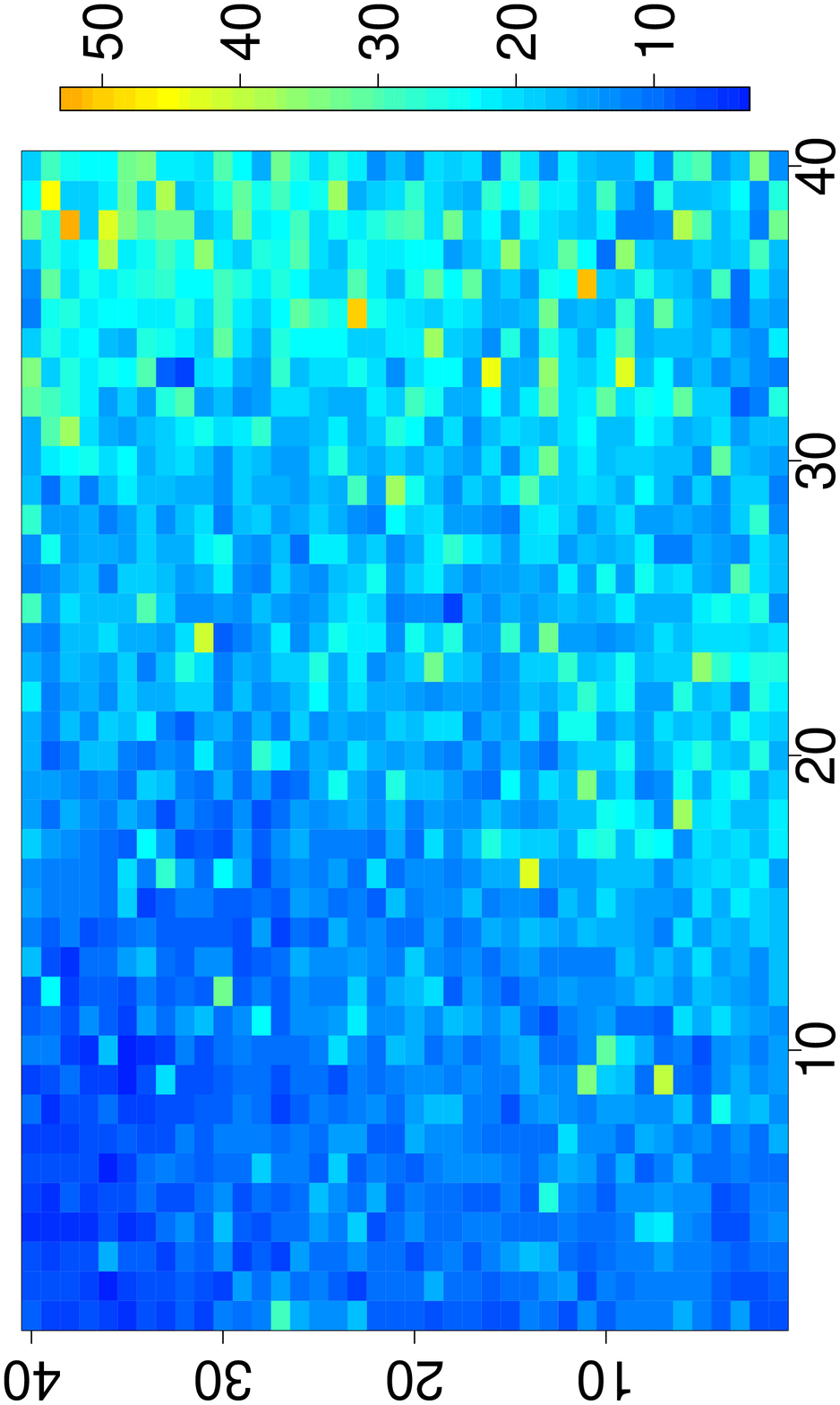}
  \includegraphics[angle=-90,width=.32\textwidth]{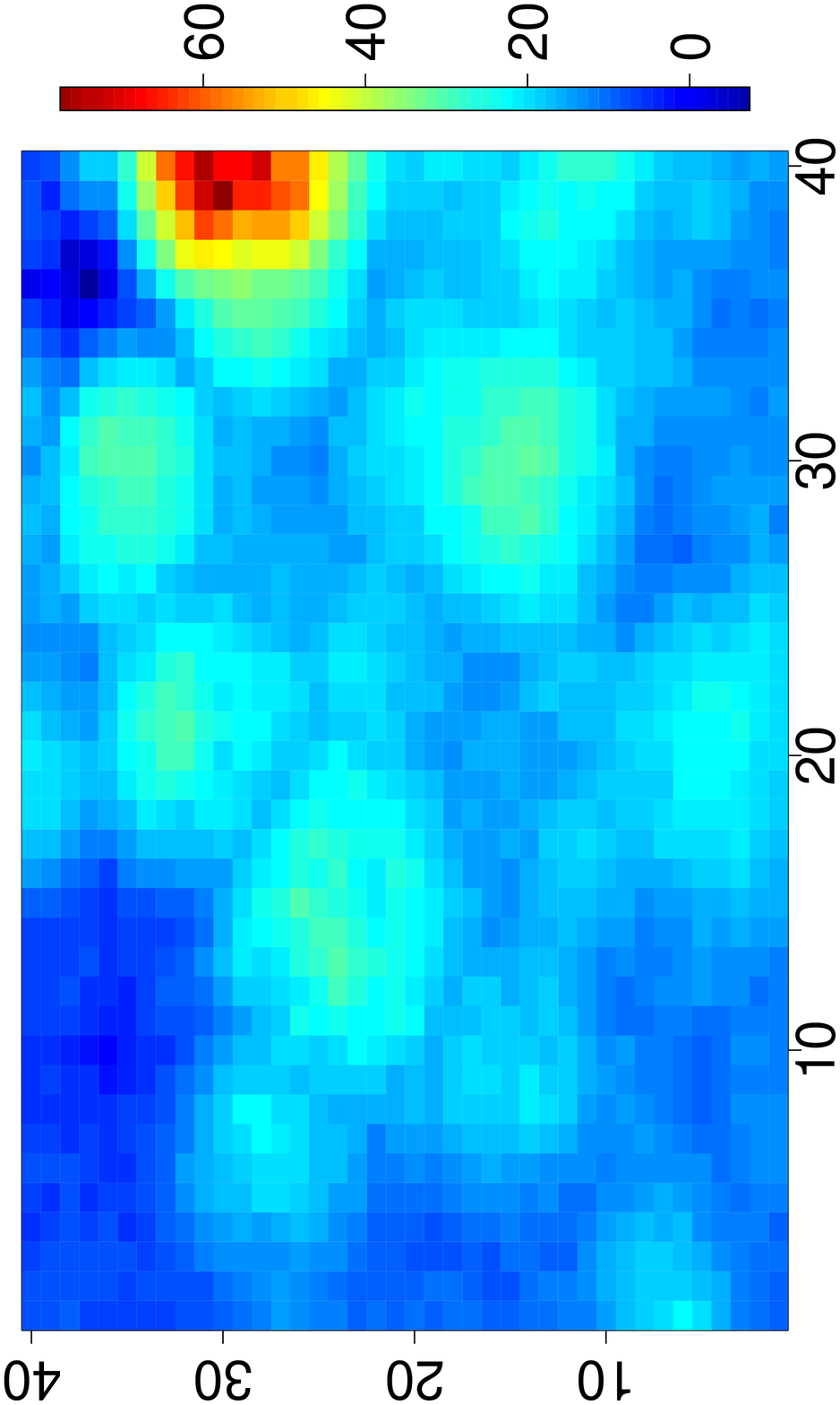}
  \caption{Comparison between one realization of the observed field
    and one realization of the different models analyzed. From left to
    right: observed field; conditional independence model; and
    max-stable hierarchical model with adjustment. The same seed was
    used for each simulation.}
  \label{fig:oneRealisation}
\end{figure}

The advantage of the max-stable hierarchical model over the conditional
independence model is that the former can account for local
dependence; even with only 50 locations in the region, it seems to
be able to detect the true pattern of local dependence.  The 95\%
credible intervals for the elements $\sigma_{11}$, $\sigma_{12}$ and
$\sigma_{22}$ of $\Sigma$ are $(5.39, 8.76)$, $(-1.28, 0.67)$ and
$(5.58, 8.37)$, which include the true values 6, 0, and 6. The fitted
max-stable model provides a mechanism for producing realistic draws
from the spatial process.  As Figure~\ref{fig:oneRealisation} shows, a
draw from the posterior distribution of the conditional independence
model would be inappropriate and unrealistic for spatial phenomena
such as rainfall or temperature, annual maxima of which would produce smoother
surfaces.

These results are obtained from a (near) perfect model simulation;
that is, the max-stable hierarchical model fitted to the data was
nearly identical to that from which the data were
simulated. Nevertheless, this simulation exercise shows that the
adjusted max-stable hierarchical model can flexibly model
marginal behavior that captures regional spatial effects and can 
capture local dependence through the max-stable process model.
Despite the approximation due to employing a composite likelihood, the
inference obtained appropriately captures the uncertainty associated
with the estimation. In the next section we show that it also seems to
perform well on real data.

\section{Application}
\label{sec:application}

\begin{figure}
  \centering
  \includegraphics[angle=-90,width=.75\textwidth]{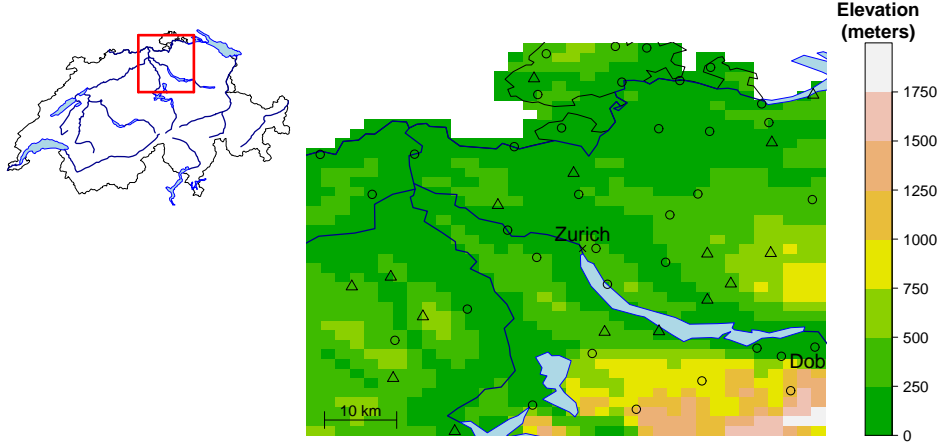}
  \caption{Map of the study region. The stations used for
    inference/validation are depicted by circles/triangles.}
  \label{fig:regionStudy}
\end{figure}

We analyze data on maximum daily rainfall amounts for the years
1962--2008 at 51 sites in the Plateau region of Switzerland; see
Figure~\ref{fig:regionStudy}. The area under study is relatively flat,
the altitudes of the sites varying from 322 to 910 meters above mean
sea level. Data from 16 of the stations
were were kept aside for model
validation and not used for fitting.

\begin{figure}
  \centering
  \includegraphics[angle=-90,width=\textwidth]{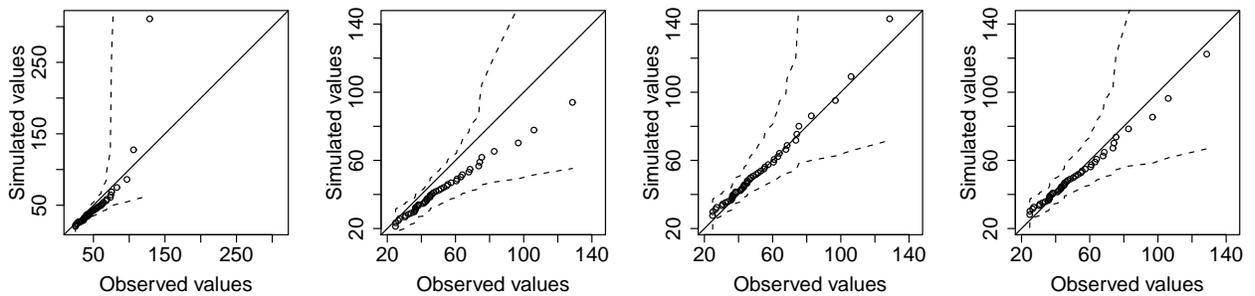}
  \caption{QQ-plots to compare the observed maxima of the annual maxima
    from the validation locations and those simulated from various
    models. From left to right: simple max-stable, conditional
    independence, unadjusted Bayesian hierarchical, adjusted Bayesian
    hierarchical models. The $95\%$ confidence/credible envelopes are
    shown as dashed lines.}
  \label{fig:checkKwiseMaxApp}
\end{figure}

Figure~\ref{fig:checkKwiseMaxApp} compares the annual maxima over the
$16$ validation stations, which we term the ``groupwise maxima'', and
the simulated groupwise maxima from the different models. All the
max-stable based models seem able to model the distributions of
the groupwise maxima, though the simple max-stable model badly
overestimates the largest value, perhaps due to inaccurate trend
surfaces for the GEV parameters, particularly the shape
parameter. The conditional independence model shows systematic
underestimation, confirming that this model is
inappropriate. The unadjusted and adjusted Bayesian hierarchical
models yield similar credible envelopes, which seem
principally to reflect the variability of simulated conditional
Gaussian processes and GEV realizations.

\begin{figure}
  \centering
  \includegraphics[angle=-90,width=\textwidth]{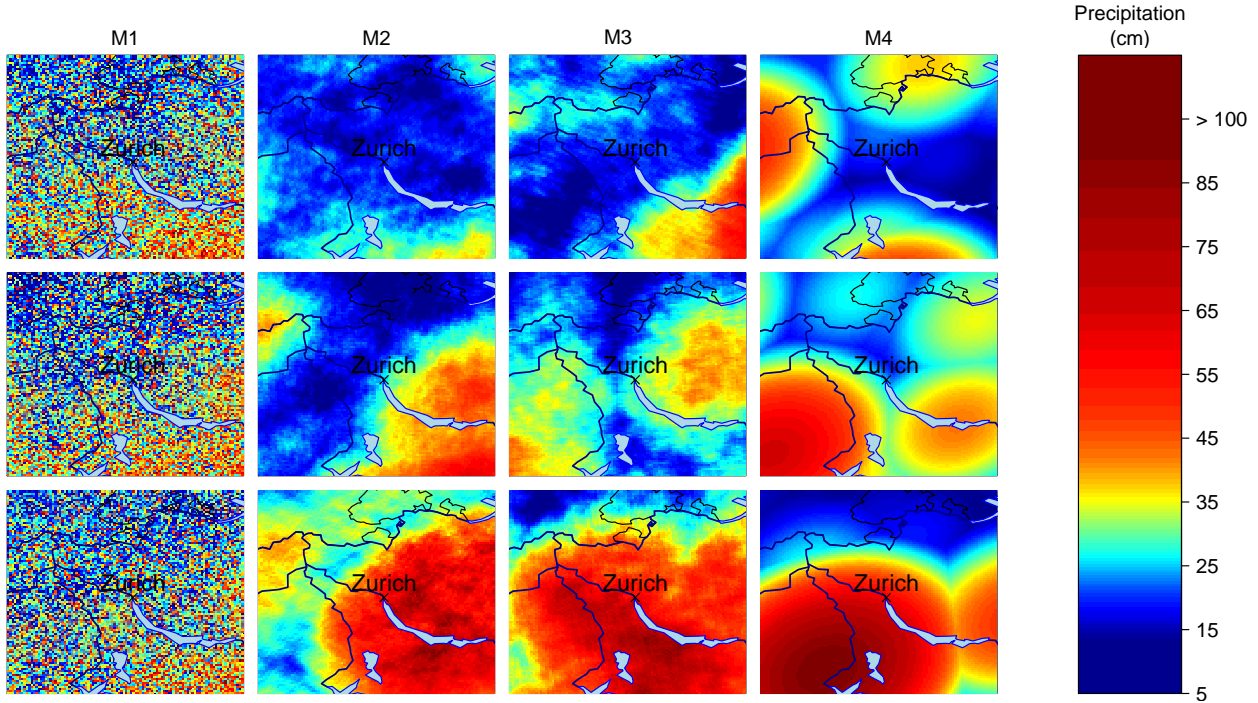}
  \caption{Three realizations of random fields over the study region
    for the conditional independent model (M1), hierarchical models
    without any adjustment (M2) and with the curvature adjustment (M3)
    and a simple max-stable model with deterministic trend surfaces
    (M4). The three rows show realizations corresponding to different
    risk scenarios according to the values of $S_{\rm Zurich}$
    expected to be exceeded once every $1.05$, $2$ and $20$ years (from
    top to bottom).}
  \label{fig:simFieldsApp}
\end{figure}

Figure~\ref{fig:simFieldsApp} shows three simulated random fields for
each model, taken from a large number of such fields. To rank these we took a disk $V_{\rm Zurich}$  of radius $6$ km and centered near the Zurich gauging station, and ordered the random fields according to
their suprema $S_{\rm Zurich}=\sup_{x \in V_{\rm
    Zurich}} Y(x)$. This allows us to summarize the intensity of a
particular realization of a random field. The three rows of
Figure~\ref{fig:simFieldsApp} correspond to the situation where
$\Pr[S_{\rm Zurich} \leq z_{\rm crit}] = \alpha$, where $\alpha =
0.05, 0.50, 0.95$ respectively and the level $z_{\rm crit}$ depends on
the model considered. Roughly speaking, the three rows show patterns
for which $S_{\rm Zurich}$ is expected to be exceeded once every $1.05$,
$2$ and $20$ years.

The conditional independence model leads to unrealistic realizations
of extreme rainfall fields, but because of the deterministic trend
surfaces for the marginal parameters, the simple max-stable model
produces fields that are too smooth to be realistic. The unadjusted
and adjusted hierarchical models seem to produce the most plausible
realizations.

\begin{figure}
  \centering
  \includegraphics[angle=-90,width=\textwidth]{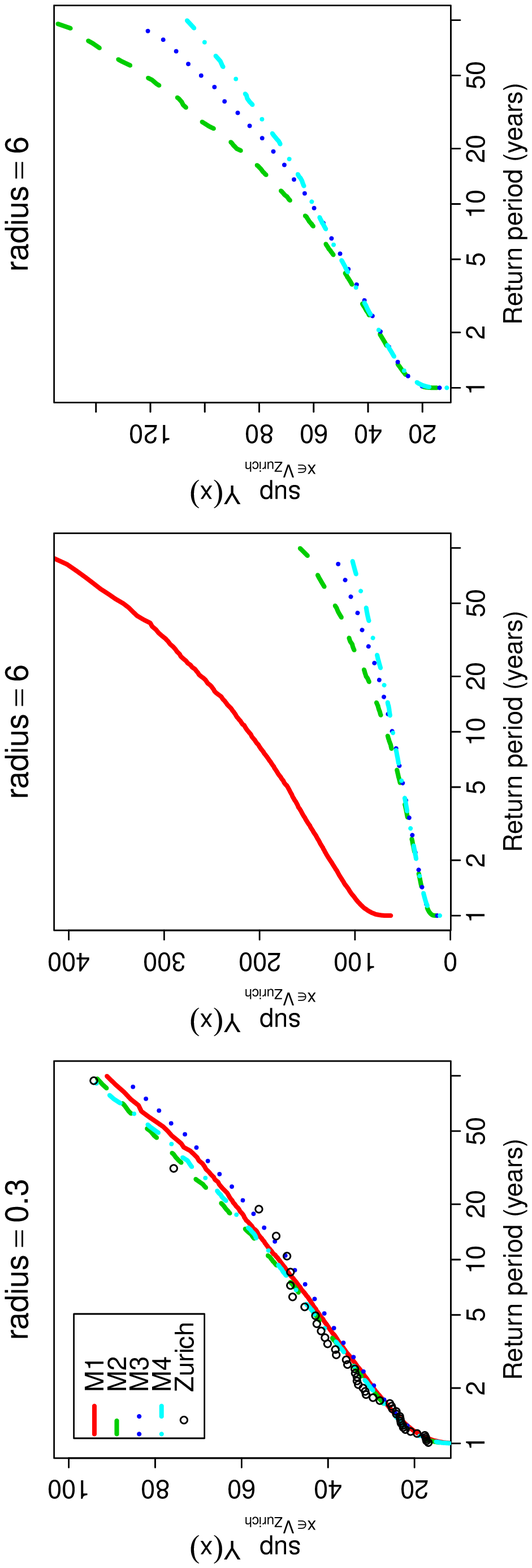}
  \includegraphics[angle=-90,width=\textwidth]{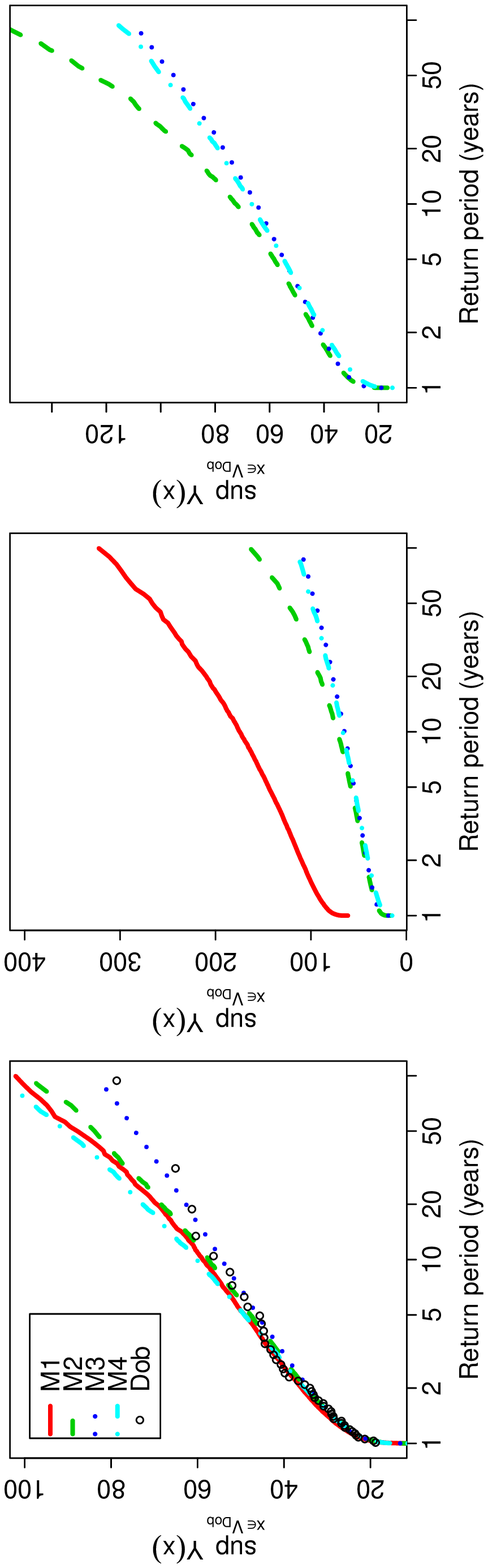}
  \caption{Comparison between the return level curves (cm) computed on
    neighborhoods centered at the Zurich (top) and DOB gauging stations
    (bottom) and having radius $0.3$ and $6$ km (left and middle
    panels) for the conditional independent model (M1), the
    hierarchical models without any adjustment (M2) and with the
    curvature adjustment (M3) and a simple max-stable model with
    deterministic trend surfaces (M4). The left panels compares the
    return level curves to the observations available at the gauging
    stations. The right panel is the same as the middle one but shows
    only the max-stable based models.}
  \label{fig:simNormApp}
\end{figure}

Figure~\ref{fig:simNormApp} plots return level curves, i.e.,\ graphs of
the estimated $p$th quantile of $S_{\rm Zurich}$ and a similar quantity $S_{\rm DOB}$
for the DOB gauging station, against $1/(1-p)$, and smaller disks of radius $0.3$. For the
smallest neighborhood, the return level curves are compared to the
observations available at the Zurich and DOB gauging stations; see
Figure~\ref{fig:regionStudy}.

As the neighbourhoods of radius 0.3km are very small, the return level
curves should be close to the empirical curves computed from the data
available at the Zurich and DOB gauging stations. This is indeed the
case for Zurich, where all the models apparently reproduce the
distribution of extreme rainfall quite well. The results are less
convincing for the DOB gauging station where, apart from the adjusted
hierarchical model, all the models seem to overestimate the largest
extremes. This situation is similar to that seen in
Section~\ref{sec:bayes-hier-model-1}: the unadjusted hierarchical
model produces a posterior that is too concentrated, while the
max-stable trend surface model might not be flexible enough. Both
models fail to capture the complicated spatial behavior of the GEV
parameters.

For the neighbourhoods of radius 6km, the central panel of the figure shows a
very strong discrepancy between the models, because of their different
spatial assumptions.  The conditional independence model yields
unrealistically high return levels, of around 2m for 10-year values,
for example.  All the max-stable based give approximately the same
return levels for return periods shorter than $10$ years. For larger
return periods, the unadjusted hierarchical model gives the largest
estimates. The same plots for 20 other gauging stations
depicted the same patterns, suggesting that the unadjusted
hierarchical model systematically overestimates the distribution of
the supremum in a given neighborhood.

\section{Conclusion}
\label{sec:conclusion}

In this paper, motivated by a real problem in which Bayesian inference
seems natural but a full likelihood is unavailable, we investigate the
usefulness of composite likelihood within a Bayesian framework. The posterior distribution obtained from a
naive implementation of a composite likelihood can have very poor
coverage properties, owing to its inappropriate re-use of the data.

To bypass this hurdle, we propose two modifications of the composite
likelihood to recover the usual asymptotic distribution of the
likelihood ratio statistic at the true value of the parameters
$\theta_0$.  We show how these adjustments can be implemented in Markov chain Monte Carlo 
algorithms and propose two ways of integrating them into the Gibbs
sampler.  Although the approximation degrades with distance
from the parameter underlying the data, simulation studies show that
the proposed framework has coverage properties similar to those
obtained using the full posterior.

The work was motivated by a need to flexibly model the marginal
distributions when modeling spatial extreme phenomena.  We construct a
Bayesian hierarchical model whose data layer is driven by a max-stable
process while the marginal parameters are modeled as realizations of a
stochastic process.  A spatial extreme simulation study showed that
this framework is able to capture complex marginal behavior as well as
the spatial dependence in the data.  An application to extreme
rainfall around Zurich shows that the approach can capture both
local dependence due to individual storms and regional dependence due
to similar climatologies, thus broadening the scope of max-stable
modelling beyond its current limits.  

\section*{Acknowledgments}
\label{sec:acknowledgements}

The work of M.~Ribatet and A.~C.~Davison was supported by the CCES
Extremes project,
\url{http://www.cces.ethz.ch/projects/hazri/EXTREMES}. D.~Cooley's
work is partly supported by National Science Foundation grant
DMS-0905315.

\appendix

\section{Asymptotic distributions of the posterior distributions}
\label{sec:asympt-distr-post}

The derivation of the asymptotic normality of the posterior
distribution heavily relies on Taylor expansions.
 Let
$\hat{\theta}_c$ denote the maximum composite likelihood estimate, let $\theta_{\rm prior}$ denote 
the mode of the prior distribution $\pi(\theta)$, and let
\begin{equation*}
  h^{\rm tot}_c(\hat{\theta}_c) = -\nabla^2_\theta \ell^{\rm tot}_c(y;\hat{\theta}_c), \qquad
  h_{\rm prior}(\theta_{\rm prior}) = -\nabla^2_\theta \log \pi(\theta_{\rm prior}).
\end{equation*}
For $n$ large enough we have 
\begin{align*}
  \pi_c(\theta    \mid   y)   &\stackrel{\cdot}{\propto}    \exp   \left\{
    \ell^{\rm tot}_c(y;\hat{\theta}_c)  -   \frac{1}{2}  (\theta  -
    \hat{\theta}_c)^T h^{\rm tot}_c(\hat{\theta}_c) (\theta - \hat{\theta}_c) +
    \log \pi(\theta_{\rm prior}) - \frac{1}{2} (\theta - \theta_{\rm
      prior})^T h_{\rm prior}(\theta_{\rm prior}) (\theta -
    \theta_{\rm prior}) \right\}\\
  &\stackrel{\cdot}{\sim} N\left\{\tilde{\theta},
    \tilde{h}(\hat{\theta}_c, \theta_{\rm prior})^{-1} \right\},
\end{align*}
where $\tilde{h}(\hat{\theta}_c, \theta_{\rm prior}) =
h^{\rm tot}_c(\hat{\theta}_c) + h_{\rm prior}(\theta_{\rm prior})$ and
$\tilde{\theta} =\tilde{h}(\hat{\theta}_c,\theta_{\rm prior})^{-1}
\{h^{\rm tot}_c(\hat{\theta}_c) \hat{\theta}_c + h_{\rm prior}(\theta_{\rm
  prior}) \theta_{\rm prior}\}$.

Provided the contribution of the prior
distribution $\pi(\theta)$ vanishes as $n \to \infty$, 
the strong law of large numbers implies that
\begin{align*}
  n^{-1} \tilde{h}(\hat{\theta}_c, \theta_{\rm prior}) = \left\{
    \frac{h^{\rm tot}_c(\hat{\theta}_c)}{n} + \frac{h_{\rm prior}(\theta_{\rm
        prior})}{n} \right\} &\longrightarrow -\mathbb{E}[\nabla^2
  \ell_c(\theta_0;Y)] = H(\theta_{0}),\\
  \tilde{\theta} = \left\{ \frac{\tilde{h}( \hat{\theta}_c,
      \theta_{\rm prior})}{n} \right\}^{-1}
  \left\{\frac{h^{\rm tot}_c(\hat{\theta}_c)}{n} \hat{\theta}_c + \frac{h_{\rm
        prior}(\theta_{\rm prior})}{n} \theta_{\rm prior}\right\}
  &\longrightarrow \theta_0,
\end{align*}
almost surely, and thus $\pi_c(\theta \mid y)
\stackrel{\cdot}{\sim} N\left\{\theta_0, n^{-1} H(\theta_{0})^{{-1}}
\right\}$.

The derivation of the asymptotic distribution for the magnitude
adjustment uses the same argument, with a slight modification. As
$n \to \infty$,
\begin{equation*}
  \hat{k} \longrightarrow p / \mbox{tr}\left\{ H(\theta_{0})^{-1}
    J(\theta_{0}) \right\} 
\end{equation*}
almost surely. Since $\hat{k}$ is estimated prior to running the MCMC
algorithm, we can assume that $\hat{k}$ is a
(tuning) constant that does not depend on $\theta$. Therefore the analogue of $h_{c}(\hat{\theta}_{c})$ when
using $\ell_{\rm magn}$ in place of $\ell^{\rm tot}_{c}$ is
\begin{equation*}
  h_{\rm magn}(\hat{\theta}_{c}) = - \hat{k} \nabla^{2}_{\theta}
  \ell_{\rm magn}(y;\hat{\theta}_{c}) \longrightarrow
  \mbox{tr}\{H(\theta_0)^{-1} J(\theta_0) \} H(\theta_0), \qquad n \to
  \infty,
\end{equation*}
almost surely, from which we conclude that $\pi_{\rm magn}(\theta \mid
y) \stackrel{\cdot}{\sim} N\left\{\theta_{0}, (n p)^{-1}
  \mbox{tr}\{H(\theta_{0})^{-1} J(\theta_{0}) \} H(\theta_{0})^{-1}
\right\}$.

We conclude with the derivation of the asymptotic distribution
of the curvature adjusted composite likelihood. By construction we have
\begin{equation*}
 n^{-1} h_{\rm curv}(\hat \theta_c) = -n^{-1}\nabla^2_\theta \ell_{\rm curv}(y;
  \hat \theta_c) \longrightarrow  H(\theta_0) J(\theta_0)^{-1}
  H(\theta_0), \qquad n \to \infty,
\end{equation*}
almost surely from which we get that $\pi_{\rm curv}(\theta \mid y)
\stackrel{\cdot}{\sim} N\left\{\theta_{0}, n^{-1} H(\theta_{0})^{-1}
  J(\theta_{0}) H(\theta_{0})^{-1} \right\}.$

\section{Asymptotic variance inflation}
\label{sec:asympt-vari-infl}

\def\pr{{\rm pr}}

In this appendix we argue that in many cases in which the densities appearing in the composite likelihood are correct, so that they satisfy the first two Bartlett identities, $\mathbb{E}[\nabla \log f(Y \in \mathcal{A}_{i} ; \theta_{0})
  ] = 0$ and $\mathbb{E}[\nabla^2 \log f(Y \in \mathcal{A}_i ;
  \theta_0)] + \mbox{Var}[\nabla \log f(Y \in \mathcal{A}_i ;
  \theta_0)] = 0$ for all $i \in I$, then   $\mbox{tr}\{H(\theta_{0})^{-1} J(\theta_{0})\} \geq p=\dim(\theta_{0})$.  This agrees with our empirical experience, which is that in many cases $\mbox{tr}\{H(\theta_{0})^{-1} J(\theta_{0})\} \gg p$.

  We first note that 
    \begin{equation*}
    \mbox{tr}\{H(\theta_{0})^{-1} J(\theta_{0})\} - p =
    \mbox{tr}\{H(\theta_{0})^{-1} J(\theta_{0}) - \mbox{Id}_{p}\} =
    \mbox{tr}[H(\theta_{0})^{-1} \{J(\theta_{0}) - H(\theta_{0})\} ]
    \geq 0.
  \end{equation*}
  Since $H(\theta_{0})^{-1}$ is positive semi-definite, the result follows if $J(\theta_{0}) - H(\theta_{0})$ is positive semi-definite, because $\mbox{tr}\{AB\} \geq 0$ when both $A$ and $B$ are positive semi-definite.

 On the one hand we have 
  \begin{equation*}
    H(\theta_{0}) = -\mathbb{E}\left[\nabla^{2} \sum_{i \in I} \log
      f(Y \in \mathcal{A}_{i}; \theta_{0}) \right] = -\sum_{i \in I}
    \mathbb{E}\left[\nabla^{2} \log f(Y \in \mathcal{A}_{i};
      \theta_{0}) \right] = \sum_{i \in I} \mbox{Var}\left[\nabla \log
      f(Y \in \mathcal{A}_{i}; \theta_{0}) \right],
  \end{equation*}
because the variance of the score equals the Fisher information for each individual summand. On the other hand we have 
  \begin{align*}
    J(\theta_{0}) &= \mbox{Var}\left[\sum_{i \in I} \nabla \log f(Y \in
      \mathcal{A}_{i}; \theta_{0}) \right]\\
    &= \sum_{i \in I} \mbox{Var}\left[\nabla \log f(Y \in \mathcal{A}_{i};
      \theta_{0}) \right] + \sum_{i,j \in I, i \neq j}
    \mathbb{E}\left[ \nabla \log f(Y \in \mathcal{A}_{i}; \theta_{0})
      \nabla \log f(Y \in \mathcal{A}_{j}; \theta_{0})^{T} \right].
  \end{align*}
Thus
  \begin{equation*}
    J(\theta_{0}) - H(\theta_{0}) = \sum_{i,j \in I, i \neq j}
    \mathbb{E}\left[ \nabla \log f(Y \in \mathcal{A}_{i}; \theta_{0})
      \nabla \log f(Y \in \mathcal{A}_{j}; \theta_{0})^{T} \right] = \sum_{i,j \in I, i < j} \mathbb{E}( U_iU_j^T + U_jU_i^T), 
  \end{equation*}
say; clearly these expectations are symmetric. To see that they will often be positive definite, let $A_i$ and $A_j$ correspond to the events $Y \in \mathcal{A}_{i}$ and $Y \in \mathcal{A}_{j}$.  If these events are independent, then $\mathbb{E}( U_iU_j^T)=0$, but if not, suppose that that we may write let $A_i = A_i'\cap A_{ij}$, $A_j = A_j'\cap A_{ij}$, for some event $A_{ij}$ such that $A_i'$ and $A_j'$ are independent conditional on $A_{ij}$.  This arises if, for example, in a Markov chain $Y \in \mathcal{A}_{i}$ corresponds to $\{Y_1=y_1,Y_2=y_2\}$, $Y \in \mathcal{A}_{j}$ corresponds to $\{Y_2=y_2,Y_3=y_3\}$, and we take $A_i'\equiv\{Y_1=y_1\}$, $A_{ij}\equiv\{Y_2=y_2\}$ and $A_j'\equiv\{Y_3=y_3\}$.  If we write $\pr(A_i) = \pr(A_i'\mid A_{ij})\pr(A_{ij})$, then the corresponding log likelihood derivative may be written as $U_i = U_i' + U_{ij}$ in a natural notation, and 
$$
\mathbb{E}( U_iU_j^T) = \mathbb{E}\{(U'_i+U_{ij})(U'_j+U_{ij})^T\} = \mathbb{E}( U'_iU_j^{'T}) + \mbox{Var}(U_{ij})= \mathbb{E}\{\mbox{Cov}(U_i',U_j'\mid A_{ij})\} + \mbox{Var}(U_{ij}),
$$
because the cross terms $\mathbb{E}(U_i'U_{ij})=\mathbb{E}(U_j'U_{ij})=0$, as may be seen by conditioning on $A_{ij}$.  If $U_i'$ and $U_j'$ are independent conditional on $A_{ij}$, then 
$\mathbb{E}( U_iU_j^T) =\mbox{Var}(U_{ij})$ is positive semi-definite; this would be the case in the Markov chain example mentioned above.  If they are not independent, but are sufficiently weakly correlated conditional on $A_{ij}$ that the term $\mbox{Var}(U_{ij})$ is dominant, then $\mathbb{E}( U_iU_j^T)$ will also be positive semi-definite, and hence so will be $J(\theta_{0}) - H(\theta_{0}) $.  This will be the case in typical applications of composite likelihood, as terms that correspond to dependent events $A_i, A_j$ will tend to be positively correlated, because they are proximate in space or time, or both.

\singlespacing

%\bibliographystyle{apalike}
%\bibliography{/Users/Mathieu/Documents/Biblio/biblio_ribatet}
\end{document}